\documentclass[twocolumn,pra,letterpaper,groupedaddress]{revtex4}

\usepackage{times,amsmath,amsfonts,amssymb,latexsym,textcomp}
\usepackage{color}
\usepackage{graphicx}
\usepackage{multirow}
\usepackage{bbm}
\usepackage{tabularx}
\usepackage{amsthm}
\usepackage{dsfont}
\usepackage{placeins}

\usepackage[T1]{fontenc}
\usepackage[latin9]{inputenc}
\usepackage[english]{babel}

\usepackage{float}
\usepackage{array}

\renewcommand{\phi}{\varphi}

\renewcommand{\epsilon}{\varepsilon}

\def\ket#1{{\lvert}#1\rangle}

\begin{document}
\title{Demonstrating an element of measurement-based quantum error correction}
\author{Stefanie~Barz$^1$, Rui~Vasconcelos$^1$, Chiara~Greganti$^1$, Michael~Zwerger$^2$, Wolfgang~Dür$^2$, Hans~J.~Briegel$^{2,3}$, Philip~Walther$^1$}
\affiliation{$^1$~University of Vienna, Faculty of Physics, Boltzmanngasse 5, 1090 Vienna, Austria,\\
$^2$~Institut für Theoretische Physik, Universität Innsbruck, Technikerstr. 25, A-6020 Innsbruck, Austria,\\
$^3$~Institut für Quantenoptik und Quanteninformation der Österreichischen Akademie der Wissenschaften, Innsbruck, Austria}

%
\begin{abstract}
In measurement-based quantum computing an algorithm is performed by measurements on highly-entangled resource states. To date, several implementations were demonstrated, all of them assuming perfect noise-free environments.
Here we consider measurement-based information processing in the presence of noise and demonstrate quantum error detection.
We implement the protocol using a four-qubit photonic cluster state, where we first encode a general qubit non-locally such that phase errors can be detected. We then read out the error syndrome and analyze the output states after decoding. Our demonstration shows a building block for measurement-based quantum computing which is crucial for realistic scenarios.
\end{abstract}

\maketitle

\section{Introduction}
%
Measurement-based quantum computation (MQC) is a framework for quantum computation that offers conceptual and practical advantages as compared to the circuit model. The most prominent example of MQC is the one-way model~\cite{Raussendorf2001, Zhang2006b, Menicucci2006, Briegel2009}, where the 2D cluster state~\cite{Briegel2001} serves as a universal resource. Quantum information is processed by sequences of (adaptive) single-qubit measurements on a highly-entangled resource state which is prepared beforehand, without the need to perform coherent gates.
MQC is particularly suited for systems such as photons, where the coherent manipulation of quantum information by gates is difficult, but the preparation of entangled states is possible by some other means. The resource state preparation can even be probabilistic, without jeopardizing the deterministic character of the overall computation. 
When a measurement-based approach is applied to special-purpose quantum information processors, one finds that specific tasks can be performed with small resource states~\cite{Raussendorf2003,Zwerger2012}. In particular, any quantum circuit acting on $N$ qubits that only contains Clifford gates~\cite{Gottesman1997} can be implemented with a resource state of size $2N$, independent of the length of the circuit. What is more, ancilla particles that are at some stage of the algorithm measured in the Pauli basis do not increase the size of the resource state. 
It follows that several tasks, including entanglement purification~\cite{Zwerger2013} or quantum error correction (QEC)~\cite{Zwerger2013a}, can be done very efficiently in a measurement-based way, i.e. with resource states of minimal size. As an additional bonus, one encounters a significantly increased robustness against noise and imperfections~\cite{Zwerger2013, Zwerger2013a}. 

Several elements of MQC and QEC have been demonstrated in photonic setups~\cite{Walther2005a, Chen2007,Lu2007a, Tokunaga2008, Vallone2008a,Vallone2010, Ukai2011,Ukai2011a, Yao2012a, Su2012,Yokoyama2013}, including elementary gates with feedforward~\cite{Prevedel2007a} as well as simple quantum algorithms~\cite{Tame2007}, together with encoding quantum information in an error correction code~\cite{Zhao2004, Yao2012}. Here we demonstrate how quantum error detection including encoding, syndrome-readout, and decoding, can be done in a measurement-based fashion, thereby providing another building block for experimental MQC. We implement in a photonic experiment a two-qubit error detection code, where the state of a qubit is encoded in two further qubits such that a phase error on \textit{one} of these qubits can be detected. The code can also be viewed as a heralded error correction code, as a phase error can be corrected if it is known which particle is subjected to noise. We implement such errors, thereby also demonstrating the process of digitalization of errors,  and show the error detecting and correcting capabilities of the code by reading out the error syndrome and performing subsequent decoding. All steps in the protocol are achieved only by single-qubit measurements on a four-qubit cluster state, thanks to the fact that all required operations are of Clifford type.

\begin{figure}
\centering
\includegraphics[width=0.45\textwidth]{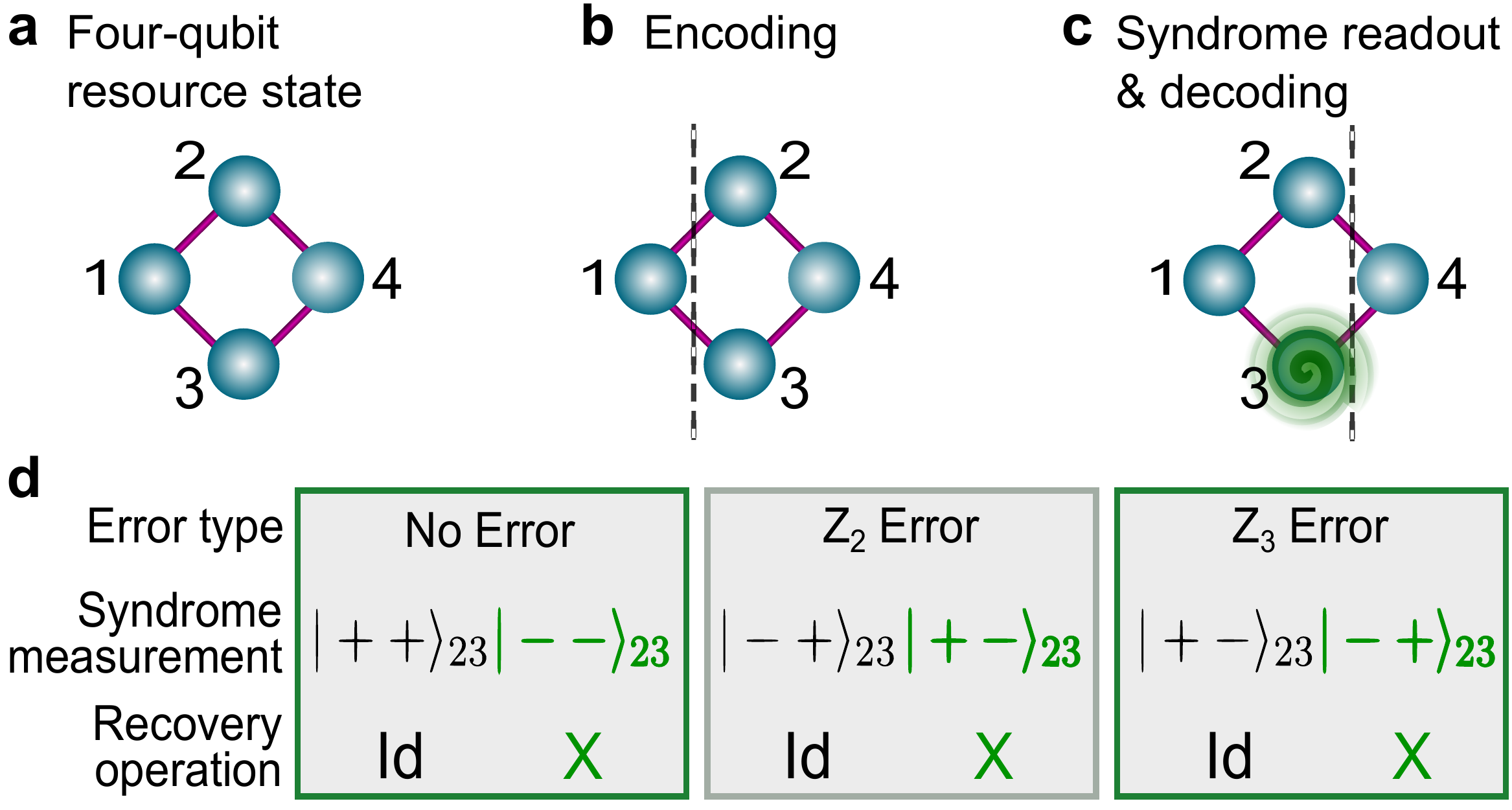}
\caption{Scheme of measurement-based error correction. \textbf{a,} The four-qubit box cluster state forms the resource. \textbf{b,} The state of qubit 1 is encoded in qubit 2 and qubit 3 by measuring it (see main text for details). \textbf{c,} An error occurs on qubit 3. \textbf{d,} Measurement instruction for the syndrome readout and respective recovery operations for different error types. Here, $Z_2$ ($Z_3$) denotes a phase error on qubit 2 (3). The result of the measurements on qubit 2 and qubit 3 shows if a $X=\sigma_x$ recovery operation needs to be applied to qubit 4. We present in the main text the analysis related to the green framed boxes, further results are shown in the Appendix.}\label{Figure1}
\end{figure}
\section{Error detection scheme}
Our protocol allows one to protect a general qubit, $\ket{\psi}= \alpha \ket{0} + \beta \ket{1}$, where $|\alpha|^2+|\beta|^2=1$, against phase noise.
The main idea of our protocol is to encode the state of a qubit $\ket{\psi}$ in two further qubits using measurement-based quantum computing~\cite{Zwerger2013a}.
By measuring these qubits, a single $Z=\sigma_Z$ error occurring on one of them can be detected, where $\sigma_Z$ is the Pauli operator.

In detail, the basis of our protocol is a four-qubit cluster state (see Figure~\ref{Figure1}), the so-called box cluster state:
\begin{eqnarray}
\label{eq:box}
 \ket{\psi_{box}}= &&\frac{1}{2}[\ket{0}_{1}\ket{+}_{2}\ket{+}_{3}\ket{0}_{4}+\ket{0}_{1}\ket{-}_{2}\ket{-}_{3}\ket{1}_{4}\\\nonumber +&&\ket{1}_{1}\ket{-}_{2}\ket{-}_{3}\ket{0}_{4}+\ket{1}_{1}\ket{+}_{2}\ket{+}_{3}\ket{1}_{4}]
\end{eqnarray}
where $\ket{\pm}=(\ket{0}\pm\ket{1})/\sqrt{2}$ are the eigenstates of the Pauli operator $\sigma_x=X$.\\
The steps of our protocol are then encoding, measurement of the error syndrome and decoding.

\begin{figure}
\centering
\includegraphics[width=0.49\textwidth]{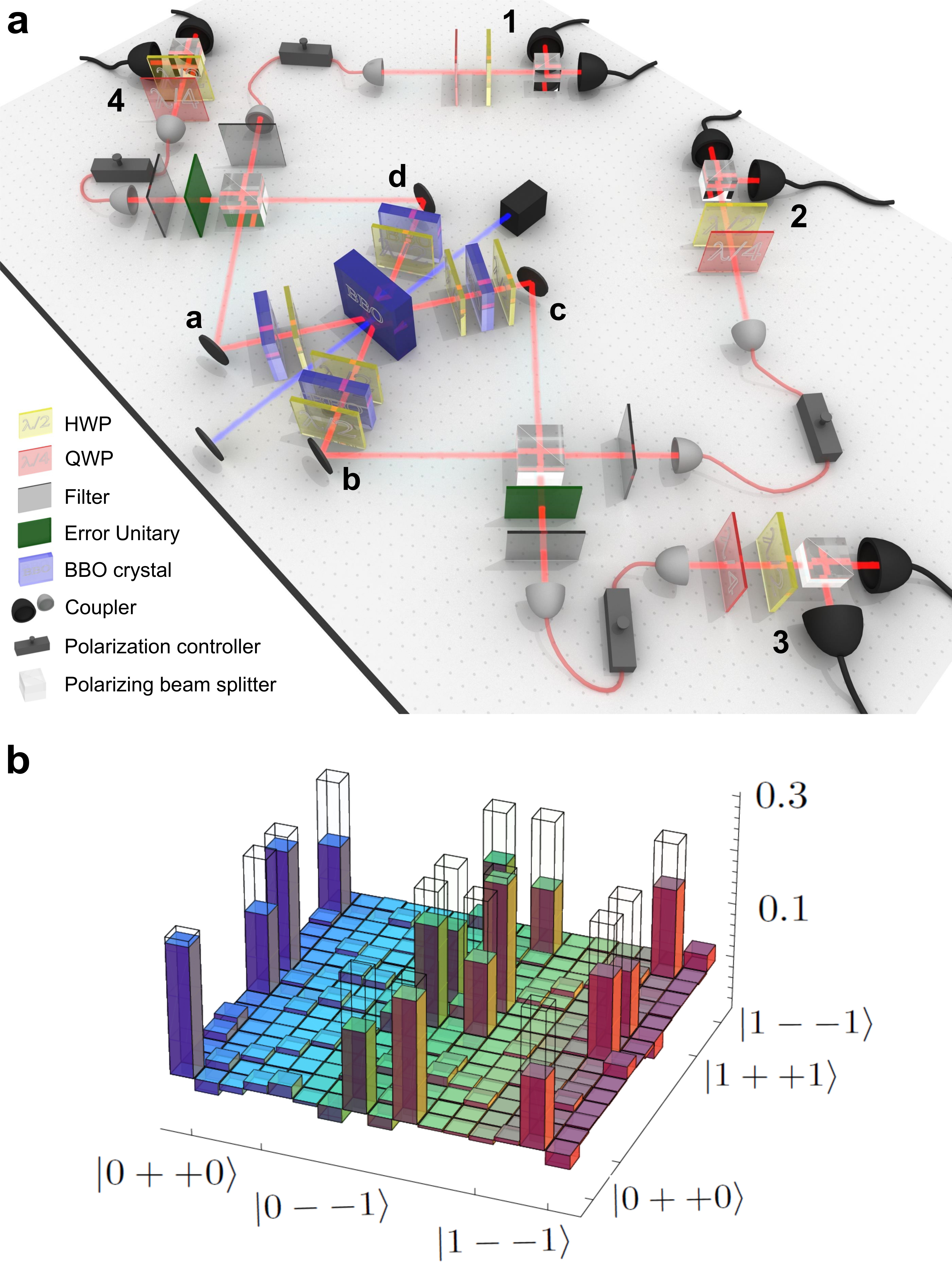}
\caption{Experimental setup and results. \textbf{a,}  A UV pump beam makes two passes through a BBO crystal, generating entangled photon pairs in the forward and backward modes. The coherent overlap of the different emissions at the polarizing beam splitters (PBSs) together with post-selection of four-fold coincidences yield the four components of the cluster state. The error unitaries, half-wave plates (HWPs) or quarter-wave plates (QWPs), implement physically the $e^{-i (\pi/2) Z}$ ($e^{-i (\pi/4) Z}$) error. Finally, the state is analyzed via the state quantum tomography using HWPs, QWPs and PBSs. \textbf{b,} Reconstructed density matrix (real part) of the four-qubit box cluster state in the eigenbasis of $Z\otimes X\otimes X\otimes Z$. The wire frame shows the ideal density matrix; the components of the imaginary part are below $0.037$ and are hence not presented here.}\label{Figure2} 
\end{figure}

First, the encoding is accomplished by a single-qubit measurement on qubit~1  in the basis $\{\alpha^*\ket{0}+\beta^* \ket{1},\beta \ket{0}-\alpha\ket{1}\}$ (see Figure~\ref{Figure1}).
If qubit~1 is projected onto the state $\alpha^*\ket{0}+\beta^* \ket{1}$, the state $\ket{\psi}$ is encoded on qubit $2$ and qubit $3$
and the remaining three-qubit state can then be written as
\begin{eqnarray}
\label{eq:3qubitstate}
 \ket{\psi_{3}}= &&\frac{\alpha}{\sqrt{2}}(\ket{++}_{23}\ket{0}_4 + \ket{--}_{23}\ket{1}_4)\\\nonumber
             +&&\frac{\beta}{\sqrt{2}}(\ket{--}_{23}\ket{0}_4 + \ket{++}_{23}\ket{1}_4).
\end{eqnarray}
In the case of the other projection, $\beta \ket{0}-\alpha\ket{1}$, the desired encoding can still be achieved as long as this state differs from $\alpha^*|0\rangle + \beta^*|1\rangle$ only by local Pauli operations, which is e.g. the case if the coefficients $\alpha$ and $\beta$ are real (see Appendix for details). 
Thus, in this example, qubits with real coefficients can be deterministically encoded, while for complex coefficients the encoding works only probabilistically. A deterministic encoding of an unknown qubit can be achieved by coupling this additional qubit by means of a Bell measurement to our resource state.

\begin{figure}
\centering
\includegraphics[width=0.45\textwidth]{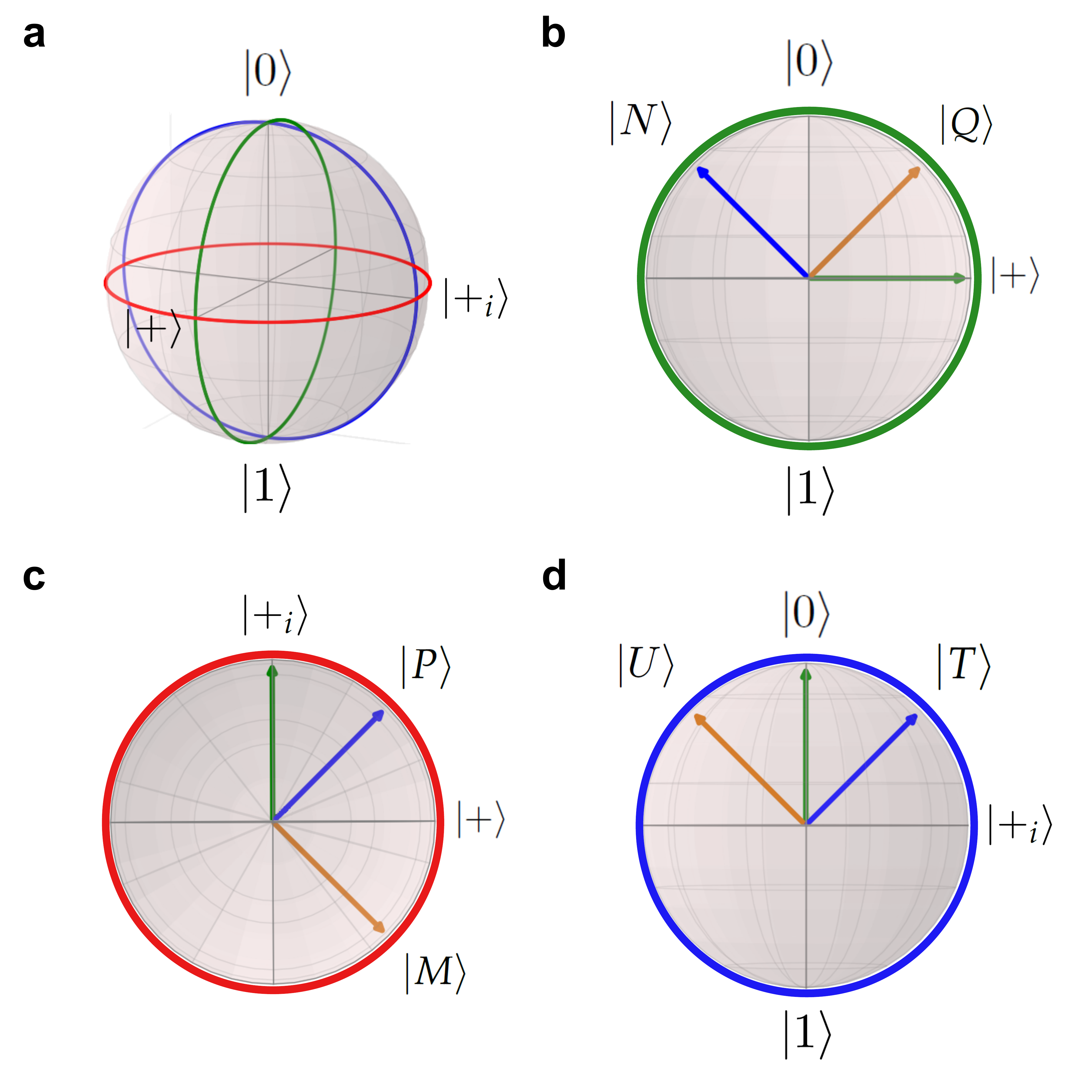}
\caption{Representation of set of encoded input states. \textbf{a,} A Bloch sphere where different planes are marked with color code. The green circle marks the X-Z plane, the red circle the X-Y plane, the blue circle the Y-Z plane. \textbf{b, c, d,} The encoded input states are shown in the correspondent  plane of the Bloch sphere. For a complete definition of the chosen states, see Table~III in the Appendix.}\label{Figure3}
\end{figure}
\begin{figure*}
\centering
\includegraphics[width=0.75\textwidth]{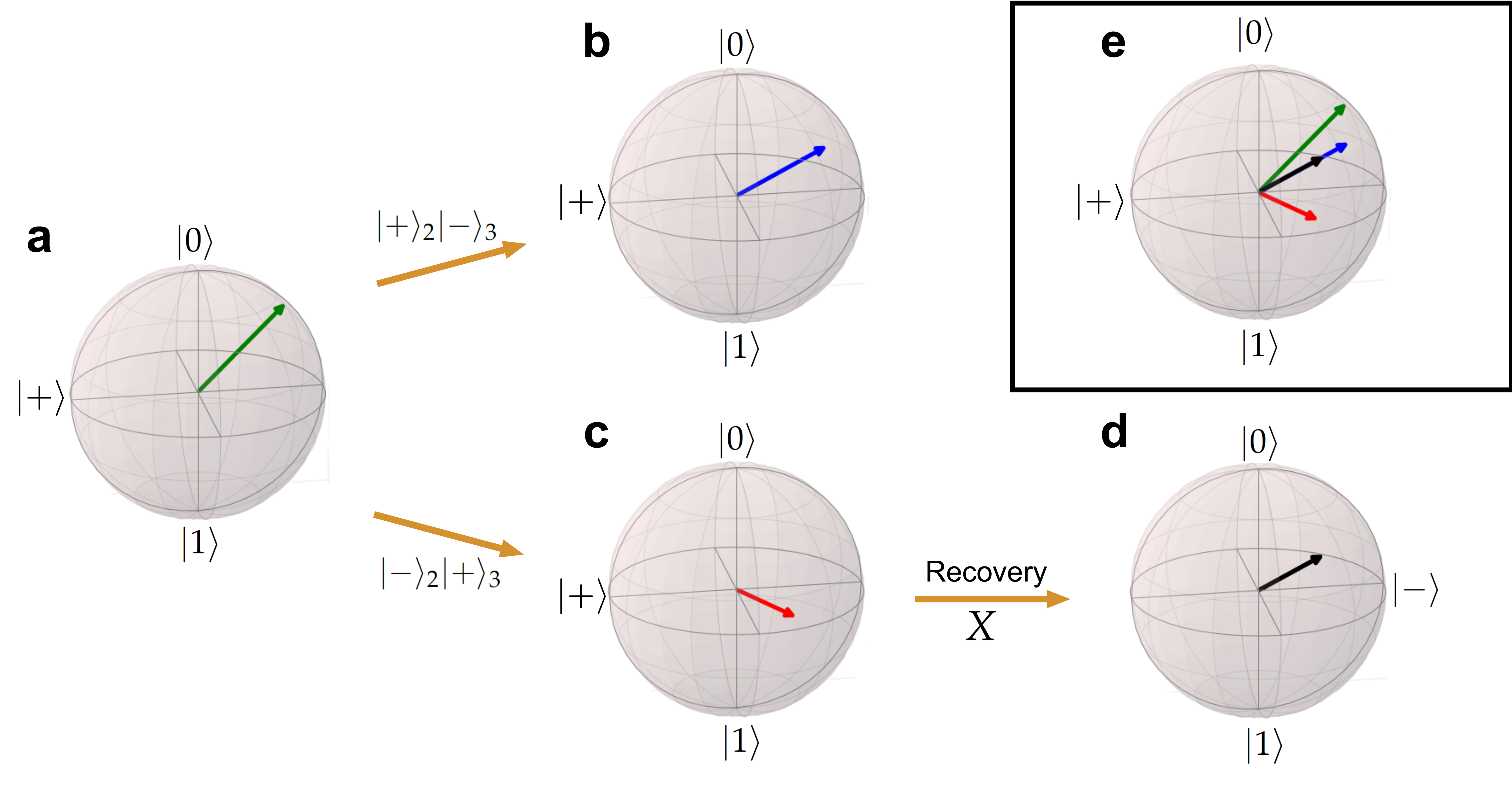}
\caption{Experimental results. We encode a state into qubits 2 and 3 and let then a phase error act on qubit 3. \textbf{a,} Representation of the ideal state $\ket{N}= (\ket{+_i}+ e^{-i \frac{\pi}{4}}\ket{-_i})/\sqrt{2}$ to be encoded ($\ket{\pm_i}=(\ket{0}\pm i\ket{1})/\sqrt{2}$). \textbf{b,} State of qubit 4 after decoding when qubits 2 and 3 are measured in the state $\ket{+-}_{23}$. In this case where no recovery operation is required (Fig.~\ref{Figure4}b) the fidelity of the decoded state is $F=0.940\pm0.024$. \textbf{c,} State of qubit 4 after decoding when qubits 2 and 3 are measured in the state $\ket{-+}_{23}$ before a recovery operation was applied. \textbf{d,} Qubit 4 after applying the recovery operation $X$ in (c) through post processing ($F=0.875\pm0.029$). \textbf{e,} Overview of the encoded and decoded qubits of (a-d).}\label{Figure4}
\end{figure*}

Now, an error can occur either on qubit~2 or qubit~3. As our protocol detects errors on one of the qubits, we assume in the following that an error occurs on qubit 3; the analysis for an error on qubit~2 is similar and hence not shown here.

The second step is to read out the error syndrome.
The protocol presented here enables the detection of an error coming from any (continuous) rotation around the Z axis.
Here, we focus on the demonstration of an $e^{-i (\pi/2)\,Z}$ error, a full phase error, or an $e^{-i (\pi/4)\,Z}$ error, which demonstrates the digitalization of errors. In the latter case, the syndrome measurement projects the coherent superposition of error and no error onto one of these possibilities. This demonstrates a crucial ingredient of quantum error correction which ensures that quantum error correcting codes can cope with continuous errors, by mapping them probabilistically to a discrete set of Pauli errors. 
The error syndrome and the decoding are achieved in a single step by measuring qubit~2 and~3 in the basis $X$.
For an error on qubit~3, if qubits~2 and~3 are found in the state $\ket{++}_{23}$ or $\ket{+-}_{23}$,  qubit~4 is in the state $\ket{\psi}$
and hence no recovery operation is necessary; if they are measured in the state $\ket{-+}_{23}$ or $\ket{--}_{23}$ a recovery operation on qubit 4 needs to be applied (see in Fig.~\ref{Figure1}d) the green framed boxes). In the case of $e^{-i (\pi/4) Z}$, these measurements determine whether an error occurs.

Finally, the remaining qubit~4 holds the decoded output.
Our protocol thus succeeds if the type and the location of the single error are known.
If the location of the error is unknown, only in the case where qubits~2 and~3 are measured in the state $\ket{++}_{23}$ or $\ket{--}_{23}$ can the encoded qubit be recovered. In that sense the scheme allows one to detect errors on either of the two intermediate qubits.

%
\section{Experiment and results}
In our experiment (see Fig.~\ref{Figure2}a), we generate the resource state using a photonic setup with polarization-entangled photons produced by spontaneous parametric down-conversion process (SPDC) in a \emph{railway-cross} scheme~\cite{Walther2005a}  (see Appendix for details about the experimental setup).
We obtain the box cluster state $\ket{\psi_{box}}$ from Eqn.~\ref{eq:box} by first experimentally producing a four-qubit cluster state (see Appendix):
\begin{eqnarray}
\label{psilab}
\ket{\psi_{lab}}=&&\frac{1}{2}[\ket{0}_{1}\ket{0}_{2}\ket{0}_{3}\ket{0}_{4}+\ket{0}_{1}\ket{0}_{2}\ket{1}_{3}\ket{1}_{4}\\\nonumber
+&&\ket{1}_{1}\ket{1}_{2}\ket{0}_{3}\ket{0}_{4}-\ket{1}_{1}\ket{1}_{2}\ket{1}_{3}\ket{1}_{4}].
\end{eqnarray}

We then apply local Hadamard gates $H$ on each qubit and a SWAP gate on qubits 2 and 4 of  $|\psi_{lab}\rangle$ to obtain:
\begin{equation}
\label{psibox}
\ket{\psi_{box}}= (H_1\otimes H_2 \otimes H_3 \otimes H_3) (\text{SWAP}_{24})\ket{\psi_{lab}}
\end{equation}

In our experiment, we perform the SWAP gate by interchanging the qubits physically and absorb the local Hadamard operations in the measurement basis.
In the following, we present all results in the basis of the box cluster state $\ket{\psi_{box}}$.

We characterize the experimentally obtained box cluster state using state tomography and
reconstruct its density matrix, $\rho$, see Fig.~\ref{Figure2}b.
For the case where no error was introduced a fidelity of $F=0.656\pm0.006$ was obtained after local unitary transformations.
The density matrices for the states after the implementation of errors have similar fidelities and are shown in the Appendix.
%
%
%

To demonstrate the implementation of the protocol we choose a set of
input states $\ket{\psi}$, as shown in Fig.~\ref{Figure3}, to be encoded in the
box cluster state and subsequently decoded, recovering the initial state.

We implement the errors on qubit 3 using additional half-wave and quarter-wave plates.
In detail, we use a HWP (QWP) at $45\text{\textdegree}$ ($-45\text{\textdegree}$) for the implementation of the $e^{-i(\pi/2)Z}$ ($e^{-i(\pi/4)Z}$) error (see Appendix).
We proceed with the error syndrome readout and, finally, the state of the decoded qubit is reconstructed through single-qubit tomography
of qubit 4.
In the case an error is detected, the original qubit is recovered through implementation of
a post-processing recovery operation: either $I$ or $X$.

The fidelities of the decoded qubits vary with
the encoded state.
For the cases where no recovery operation was needed
the fidelities are within the values $[0.810\pm0.036,\,0.990\pm0.009]$.
Decoded qubits, in which a $X$ recovery operation was applied,
present slightly lower fidelities, $[0.629\pm0.039,\,0.982\pm0.008]$.
This discrepancy follows from non-ideal resource states as shown in Fig.~\ref{Figure2}b.
Due to experimental noise, the single-qubit fidelities of qubit~4 vary for different projections of qubit~2 and qubit~3.

To illustrate some of the experimental results, in Fig.~\ref{Figure4} we show the results obtained for the case where the state $\ket{N}= (\ket{+_i}+ e^{-i \frac{\pi}{4}}\ket{-_i})/\sqrt{2}$ (Here, $\ket{\pm_i}=(\ket{0}\pm i\ket{1})/\sqrt{2}$) was encoded (Fig.~\ref{Figure4}), subjected to a full phase error on qubit 3, and subsequently decoded.

A list of the fidelities of all operations performed is shown in the Appendix, where we also show the results for errors occurring on qubit 2.
%
\section{Conclusion}
We have presented the implementation of an error-detection protocol in measurement-based quantum computing.
Although the demonstration was performed using a photonic quantum computing architecture, measurement-based error detection and correction can be implemented with other physical systems as demonstrated recently with trapped ions~\cite{Lanyon2013}.

Our protocol can be readily extended to larger cluster states containing more qubits.
A five-qubit cluster state would be sufficient to also correct the error within the experiment such that no post-processing would be necessary. A structure containing seven qubits or more would allow for the detection and correction of multiple phase errors, or a general error on a single qubit.

Our experiment constitutes a building block for larger-scale hybrid quantum computing networks, where elements of different computational schemes are combined
to provide a computational architecture that unifies the advantage of the different approaches~\cite{Zwerger2013a}.
In such a hybrid architecture, elementary blocks and gate sequences can be performed in a measurement-based way, i.e. by preparing specific resource states, and then
combined in a sequential fashion as in the circuit model. This approach leads to a remarkable robustness against noise and imperfections, with error thresholds on the order of 10\% per particle. Our work presents a proof-of-principle demonstration of one of the main building blocks in this scheme, thereby providing another step towards measurement-based quantum information processing in realistic scenarios.

{\bf Acknowledgments}
This work was supported by the European Commission, Q-ESSENCE (No. 248095),
QUILMI (No.295293), EQUAM (No. 323714), PICQUE (No. 608062), GRASP (No.
613024) and the ERA-Net CHISTERA project QUASAR, the John Templeton
Foundation, the Vienna Center for Quantum Science and Technology (VCQ), the
Austrian Nanoinitiative NAP Platon, the Austrian Science Fund (FWF) through
the SFB FoQuS (No. F4006-N16 and F4012-N16), START (No. Y585-N20), grant (No. P24273-N16)
and the doctoral programme CoQuS, the Vienna Science and Technology Fund
(WWTF) under grant ICT12-041, and the Air Force Office of Scientific
Research, Air Force Material Command, United States Air Force, under grant
number FA8655-11-1-3004.


\section{Appendix A: Theory}



\subsection*{Deterministic and probabilistic encoding}
As described in detail in the main paper, we encode the input state by measuring qubit~1 in the basis $\{\alpha^*\ket{0}+\beta^* \ket{1},\beta \ket{0}-\alpha\ket{1}\}$.
If qubit 1 is projected onto the state $\alpha^*|0\rangle + \beta^* |1\rangle$, the remaining three-qubit state is:
\begin{eqnarray}
\label{eq:psiOutcome1} 
 \ket{\psi_{3}}= &&\frac{\alpha}{\sqrt{2}}(\ket{++}_{23}\ket{0}_{4}+\ket{--}_{23}\ket{1}_{4})\\\nonumber
 &&+ \frac{\beta}{\sqrt{2}} (\ket{--}_{23}\ket{0}_{4}+\ket{++}_{23}\ket{1}_{4});
\end{eqnarray}
and if qubit 1 is projected onto the state $\beta|0\rangle - \alpha |1\rangle$, it is:
\begin{eqnarray}
\label{eq:psi3__} 
 \ket{\psi_{3}}=&& \frac{\beta^*}{\sqrt{2}} (\ket{++}_{23}\ket{0}_{4}+\ket{--}_{23}\ket{1}_{4})\\\nonumber
  &&- \frac{\alpha^*}{\sqrt{2}} (\ket{--}_{23}\ket{0}_{4}+\ket{++}_{23}\ket{1}_{4});
\end{eqnarray}
In the latter case, a correction is possible whenever $\beta \ket{0}-\alpha\ket{1}$ differs from $\alpha^*|0\rangle + \beta^*|1\rangle$ only by local Pauli operations. This is the case for real coefficients $\alpha,\beta$, where the desired state can be obtained by applying local Pauli corrections $( ( X Z)_2 \otimes Z_3)$, but e.g. also for a $\sigma_y$ eigenstate, $\alpha=1/\sqrt{2}, \beta =\pm i/\sqrt{2}$. In all other cases, the encoding procedure is probabilistic. Notice that the process can be made deterministic by using an additional qubit $1'$, whose (unknown) state can be read in deterministically (up to Pauli correction) by performing a Bell measurement on qubits $1,1'$.

\begin{table*}
\centering
\begin{tabular}{>{\centering}m{2.4cm}m{0.5cm}>{\centering}m{1.3cm}>{\centering}m{1.3cm}m{0.5cm}>{\centering}m{1.3cm}>{\centering}m{1.3cm}m{0.5cm}>{\centering}m{1.3cm}>{\centering\arraybackslash}m{1.3cm}}
 \multicolumn{1}{c}{Error Type}& & \multicolumn{2}{c}{No Error} & & \multicolumn{2}{c}{$Z_{2}$ Error}& & \multicolumn{2}{c}{$Z_{3}$ Error}\\
\hline
Syndrome measurement& & $\ket{++}_{23}$ & $\ket{--}_{23}$ & & $\ket{-+}_{23}$ &$\ket{+-}_{23}$  & & $\ket{+-}_{23}$ & $\ket{-+}_{23}$ \\
\hline 
\hline
Recovery operation& & $\mathds{I}$ & $X$ & &  $\mathds{I}$ & $X$ & &   $\mathds{I}$ & $X$\tabularnewline
\end{tabular}
\caption{Recovery Operations given the syndrome readout for ($e^{-i\frac{\pi}{2}Z}$) phase error.}
\label{tab:syndr1}
\end{table*}

\begin{table*}
\centering
\begin{tabular}{>{\centering}m{2.4cm}m{0.5cm}>{\centering}m{1.3cm}>{\centering}m{1.3cm}m{0.5cm}>{\centering}m{1.3cm}>{\centering}m{1.3cm}m{0.5cm}>{\centering}m{1.3cm}>{\centering}m{1.3cm}m{0.5cm}>{\centering}m{1.3cm}>{\centering\arraybackslash}m{1.3cm}}
 \multicolumn{1}{c}{} & & \multicolumn{5}{c}{ $e^{-i\frac{\pi}{4}Z_2}$ Error} && \multicolumn{5}{c}{ $e^{-i\frac{\pi}{4}Z_3}$ Error}\\
\hline
\multicolumn{1}{c}{Error Type} && \multicolumn{2}{c}{No Error}& & \multicolumn{2}{c}{$Z_{2}$ Error} && \multicolumn{2}{c}{No Error}& & \multicolumn{2}{c}{$Z_{3}$ Error}\\
\hline
Syndrome measurement& & $\ket{++}_{23}$ & $\ket{--}_{23}$ &  & $\ket{-+}_{23}$ &  $\ket{+-}_{23}$ & & $\ket{++}_{23}$ & $\ket{--}_{23}$ & & $|+-\rangle_{23}$  & $\ket{-+}_{23}$ \\
\hline 
\hline
Recovery operation & & $\mathds{I}$ & $X$ & & $\mathds{I}$ & $X$ & & $\mathds{I}$ & $X$ & & $\mathds{I}$ & $X$\\ 
\end{tabular}
\caption{Recovery Operations given the syndrome readout for ($e^{-i\frac{\pi}{4}Z}$) error.}
\label{tab:syndr2}
\end{table*}
\subsection*{Syndrome read-out \& decoding}
If a $e^{-i\frac{\pi}{2}Z}$ error occurs on qubit 2, the three-qubit state related to \eqref{eq:psiOutcome1} becomes:
\begin{eqnarray}
\label{eq:psi3error2} 
 |{\psi_{3}}^{'}\rangle= &&\frac{\alpha}{\sqrt{2}}(\ket{-+}_{23}\ket{0}_{4} +\ket{+-}_{23}\ket{1}_{4})\\\nonumber
                         &&+ \frac{\beta}{\sqrt{2}}(\ket{+-}_{23}\ket{0}_{4} +\ket{-+}_{23}\ket{1}_{4}).
\end{eqnarray}
If qubit 2 and 3 are measured to be in the state $\ket{-+}_{23}$, the final state of qubit 4 is $\ket{\psi}= \alpha \ket{0} + \beta \ket{1}$; whereas if they are measured to be in the state $\ket{+-}_{23}$, the state of qubit 4 will be $\alpha \ket{1} + \beta \ket{0}$ which can be corrected by applying a recovery operation $X$.

If a $e^{-i\frac{\pi}{2}Z}$ error occurs on qubit 3, the three-qubit state related to \eqref{eq:psiOutcome1} becomes:
\begin{eqnarray}
\label{eq:psi3error3} 
 \ket{{\psi_{3}}^{'}}= &&\frac{\alpha}{\sqrt{2}}( \ket{+-}_{23} \ket{0}_{4}  + \ket{-+}_{23} \ket{1}_{4}) \\\nonumber
                       &&+ \frac{\beta}{\sqrt{2}}( \ket{-+}_{23} \ket{0}_{4}+\ket{+-}_{23} \ket{1}_{4}).
\end{eqnarray}
In this case, if qubit 2 and 3 are projected onto the state \mbox{$\ket{+-}_{23}$}, the final state of qubit 4 will be $\ket{\psi}$, whereas for $\ket{-+}_{23}$ we obtain $\ket{\psi}$ only after applying a recovery operation $X$.

If a $e^{-i\frac{\pi}{4}Z}$ error occurs on qubit 2, the three-qubit state related to \eqref{eq:psiOutcome1} becomes:
\begin{eqnarray}
\label{eq:psi3errorHalf2} 
 |{\psi_{3}}^{'}\rangle= &&\frac{\alpha}{2}(\ket{++}_{23}\ket{0}_{4}-i\ket{-+}_{23}\ket{0}_{4}\\\nonumber
 &&+ \ket{--}_{23}\ket{1}_{4}-i\ket{+-}_{23}\ket{1}_{4}) \\\nonumber
 &+& \frac{\beta}{2}(\ket{--}_{23}\ket{0}_{4} -i\ket{+-}_{23}\ket{0}_{4}\\\nonumber
 &&+ \ket{++}_{23}\ket{1}_{4}-i\ket{-+}_{23}\ket{1}_{4}).
\end{eqnarray}
The remaining state is a coherent superposition of error case and no error case.
If qubit 2 and 3 are projected onto the state  $\ket{++}_{23}$ or  $\ket{--}_{23}$, then the encoded qubit has not been affected by noise, whereas for  $\ket{+-}_{23}$ or  $\ket{-+}_{23}$ a phase flip acted. In both cases the final state of qubit 4 will be equal to $\ket{\psi}$,  respectively  up to $\mathds{I}$ or $X$ operations.

The same procedure is used to obtain the final state of qubit 4 after a  $e^{-i\frac{\pi}{4}Z}$ error occured on qubit 3.

We summarize in the tables I and II  the syndrome outcomes and respective recovery operations for a $e^{-i\frac{\pi}{2}Z}$ error (see Table~\ref{tab:syndr1}) and a $e^{-i\frac{\pi}{4}Z}$ error  (see Table~\ref{tab:syndr2}).


\section{Appendix B: Experimental Setup}
In our experiment (see Fig.~2a in the main paper) entangled photon
pairs are produced by a non-collinear type II SPDC process on a BBO
($\beta-$Barium Borate) crystal.

A solid state $532$nm laser (Coherent Verdi-10) pumps a mode-locked
Ti:Sa oscillator (Coherent Mira 900), yielding a pulsed output ($\tau=200$fs,
$\lambda=789$nm, $76$MHz). This is afterwards frequency-doubled
through SHG in a 2mm-thick Lithium triborate (LBO) crystal, producing
UV pulses with a $0.7$W cw average.
We achieve a stable source of UV-pulses by translating the LBO to avoid optical damage to the anti-reflection coating of the crystal.
 Dichroic mirrors are used to
separate the up-converted UV from the residual infrared light.

The UV pump beam is focused on the $2\,$mm-thick BBO, generating down-converted
infrared photons in the forward modes, $a$ and $b$. Then, the UV
beam is reflected back, crossing the BBO a second time and producing
entangled photon pairs in the backward mode, $c$ and $d$.
Half-wave plates (HWPs) and additional BBOs are used to compensate for walk-off effects and allow the production of any Bell state in the forward and backward mode.

The modes of the different pairs $a$, $b$ and $d$, $c$, respectively, are then coherently overlapped at polarizing beam splitters (PBSs) by equalizing the different path lengths.

Narrow-band interference filters ($\Delta\lambda$ = $3\,$ nm) are used to spatially and spectrally select the down-converted photons which are then coupled into single-mode fibers that guide them to the polarization analysis setup. There, different polarization measurements are performed using quarter-wave plates (QWPs), HWPs and polarizing beam splitters as well as single photon detectors (Perkin Elmer - SPCM AQ4C).

The preparation of the four-qubit linear cluster state relies on the
simultaneous detection of one (and only one) photon in each of the
four outputs $1$,$2$,$3$ and $4$.

In order to produce the desired state, we align our setup such that a $\ket{{\Phi}^-}_{ab}=(\ket{HH}_{ab}-\ket{VV}_{ab})/\sqrt{2}$
state is emitted in the forward direction and a $\ket{{\Phi}^+}_{cd}=(\ket{HH}_{cd}+\ket{VV}_{cd})/\sqrt{2}$ state in the backward direction, where $\ket{H}$ ($\ket{V}$) denotes the horizontal (vertical) polarization state.
The emission of only one entangled pair in the forward direction and only one pair in the backward direction  results in two different four-photon terms:
$\ket{H}_1\ket{H}_2\ket{H}_3\ket{H}_4$ and $- \ket{V}_1\ket{V}_2\ket{V}_3\ket{V}_4$ due to the properties of the PBSs. The two-pair emissions also lead to fourfold coincidences, namely to a $-\ket{H}_1\ket{H}_2\ket{V}_3\ket{V}_4$ state and a $\ket{V}_1\ket{V}_2\ket{H}_3\ket{H}_4$ state for a double-pair emission in the forward and in the backward direction, respectively.
We shift the phase of the term
$-\ket{H}_1\ket{H}_2\ket{V}_3\ket{V}_4$ by $\pi$ to generate a sign shift.
For this, we use the method~\cite{Walther2005a} where a rotation of an additional wave plate has the desired effect. The final output state is a superposition of all these four terms.

Theoretical considerations show that for a ratio of
1:3 between the backward ($4.4\mbox{ks}^{-1})$ and forward ($13.2\mbox{ks}^{-1})$ two-fold coincidences, the right amplitudes are attained by setting the HWP to $27.\text{5\textdegree}$~\cite{Walther2005a}. The ratio is adjusted by tweaking the coupling efficiencies of the forward and backward modes.

Additional phase shifts arising due to reflections at the PBS are compensated by tilting the BBO crystals in the forward direction (and effectively aligning for a state $\ket{{\Phi}^+}$).

In our experiment, the emitted Bell pairs show typical visibilities of about $0.9$.
The different photon emissions then interfere at the PBSs with average visibilities of $0.85$.
Additional errors arise due to phase drifts during the measurements.
These main error contributions, together with minor errors like polarization drifts, decrease the fidelity of our cluster states with respect to the ideal state. In our calculations, we always assume Poissonian errors. In fact, these indicate a lower bound for the actual error that takes all the experimental imperfections into account.

The implementation of the errors is accomplished by inserting additional QWPs and HWPs in the respective modes.
Note that due to the swap operation, noise affecting qubit 2 of the box cluster is experimentally implemented on qubit 4 of our experimental cluster state. Likewise, because of the H gate, the phase flip $e^{-i\frac{\pi}{2}Z}$ is implemented via a bit flip (X) in the experiment.

The error was estimated running a 100-cycles Monte Carlo simulation
with Poissonian noise added to the experimental counts.

\section{Appendix C: Four-qubit density matrices}

We present the full-tomographic reconstructions of the box cluster resource state after the occurence of a $e^{-i\frac{\pi}{2}Z}$ error on qubit 2, a $e^{-i\frac{\pi}{2}Z}$ error on qubit 3, a $e^{-i\frac{\pi}{4}Z}$ error on qubit 2 and a $e^{-i\frac{\pi}{4}Z}$ error on qubit 3. The density matrices are presented in the eigenbasis of $Z\otimes X \otimes X \otimes Z$ to easily visualize the state. The wire frames represent
the ideal state after the error occurred.\\

\begin{figure}[H]
\centering
\includegraphics[width=0.33\textwidth]{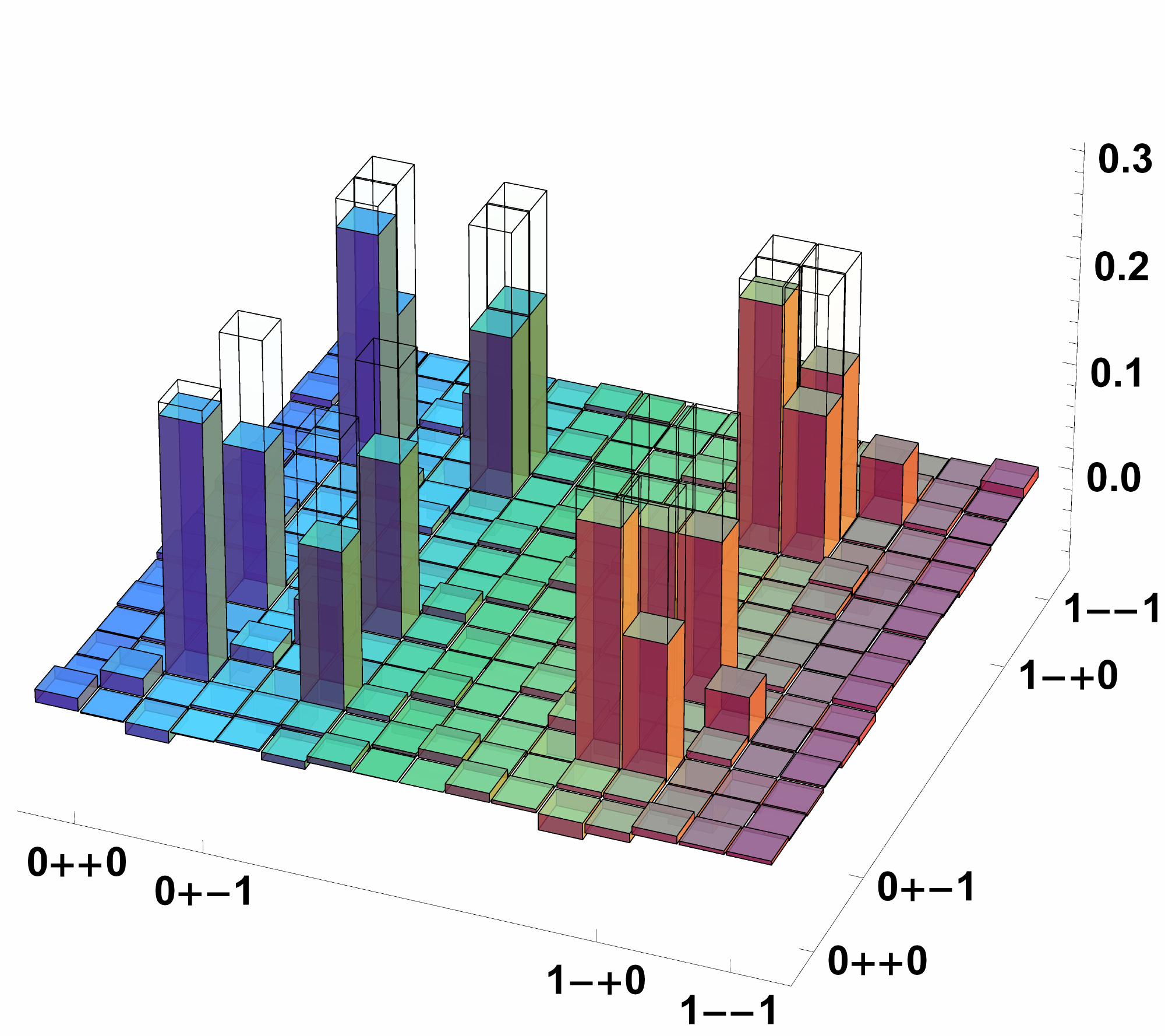}
\caption{Density matrix (real part) of the four-qubit box
cluster state after a phase error occurred on qubit 3 ($F=0.656 \pm 0.006$ via local unitary operations). 
The components of the imaginary part are below $0.032$ and are hence not presented here. }
\label{fig:PlotErrFull3}
\end{figure}

\begin{figure}[H]
\centering
\includegraphics[width=0.33\textwidth]{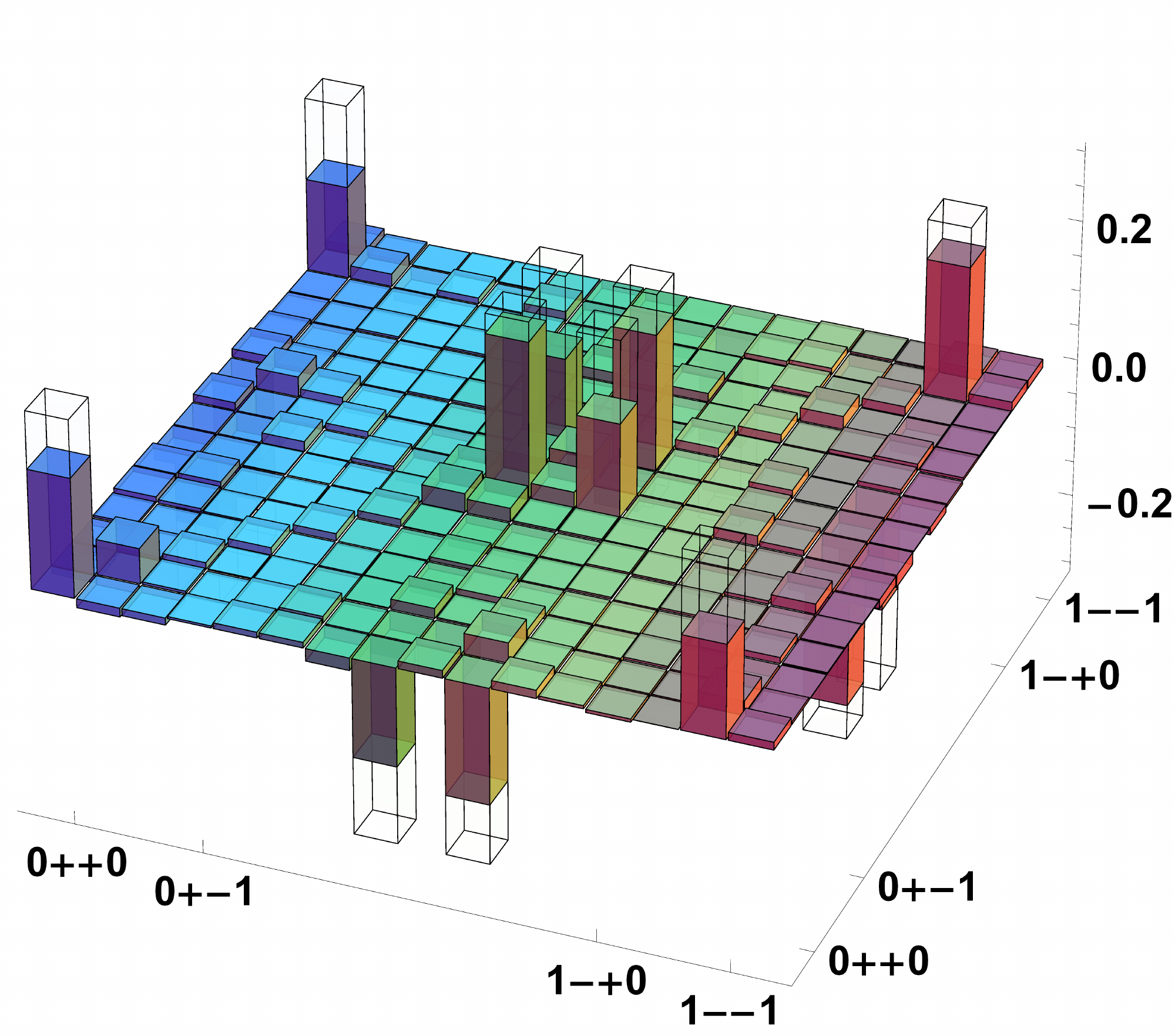}
\caption{Density matrix (real part) of the four-qubit box
cluster state after a phase error occurred on qubit 2 ($F=0.634\pm 0.008$ via local unitary operations rotations).
The components of the imaginary part are below $0.028$ and are hence not presented here. }
\label{fig:PlotErrFull2}
\end{figure}

\begin{figure}[H]
\centering
\includegraphics[width=0.33\textwidth]{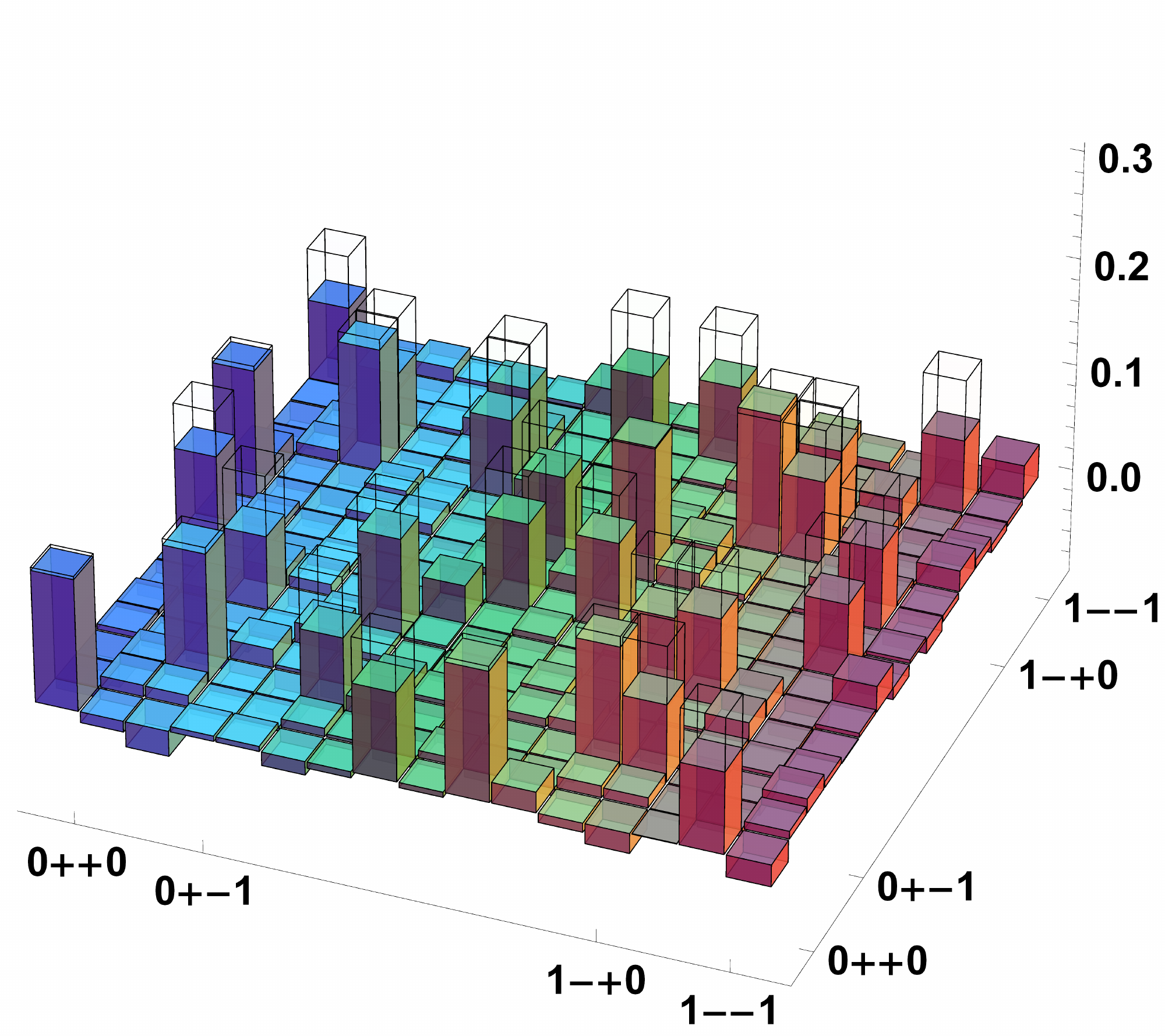}
\caption{Density matrix (real part) of the four-qubit box
cluster state after a  half phase ($e^{-i\frac{\pi}{4}Z}$) error occurred on qubit 3 ($F= 0.667 \pm 0.009 $ via local unitary operations). 
}
\label{fig:PlotErrHalf3}
\end{figure}

\begin{figure}[H]
\centering
\includegraphics[width=0.33\textwidth]{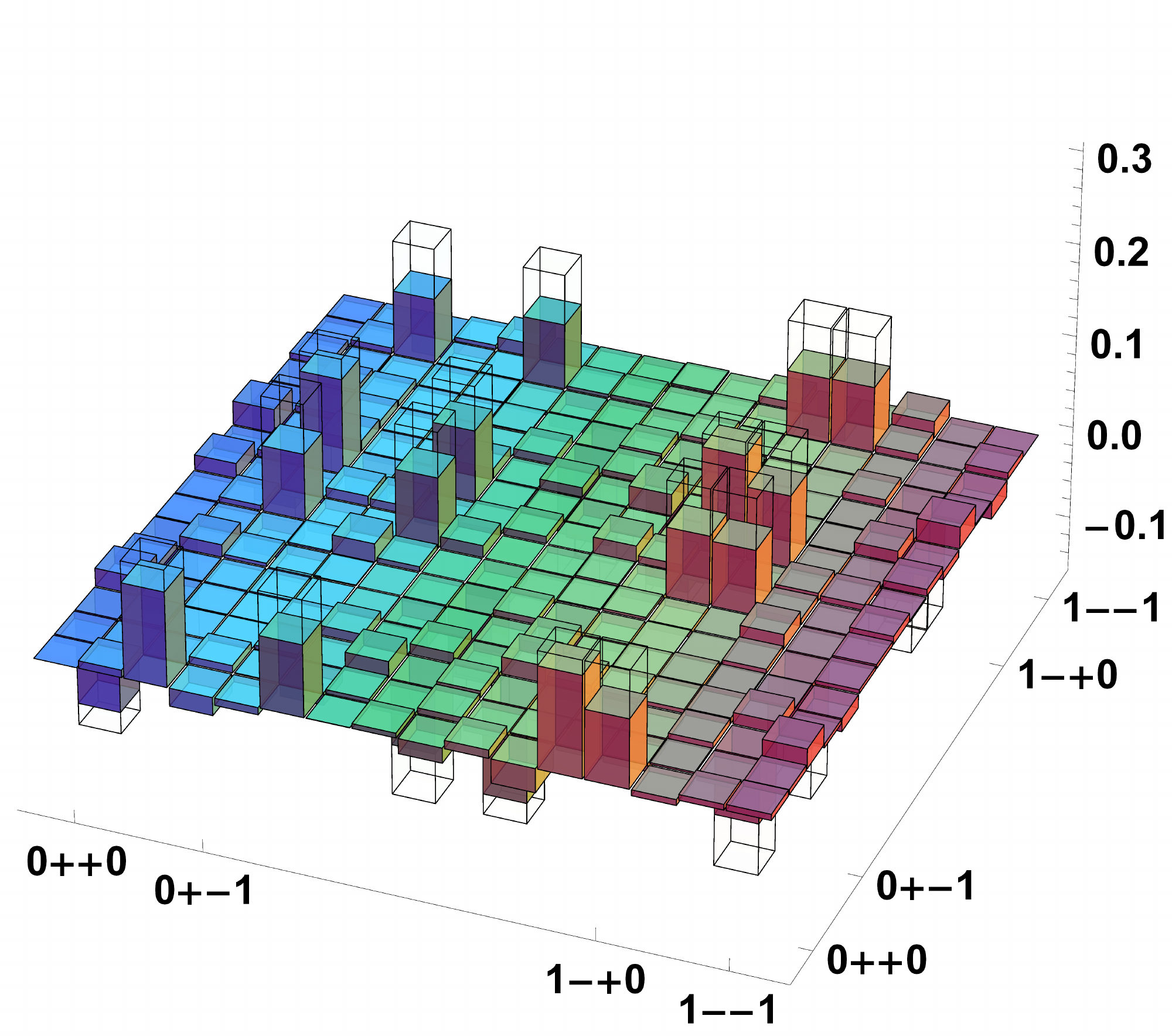}
\caption{Density matrix (imaginary part) of the four-qubit box
cluster state after a half phase ($e^{-i\frac{\pi}{4}Z}$) error occurred on qubit 3 (same fidelity as real part, Figure ~\ref{fig:PlotErrHalf3}).}
\end{figure}

\begin{figure}[H]
\centering
\includegraphics[width=0.33\textwidth]{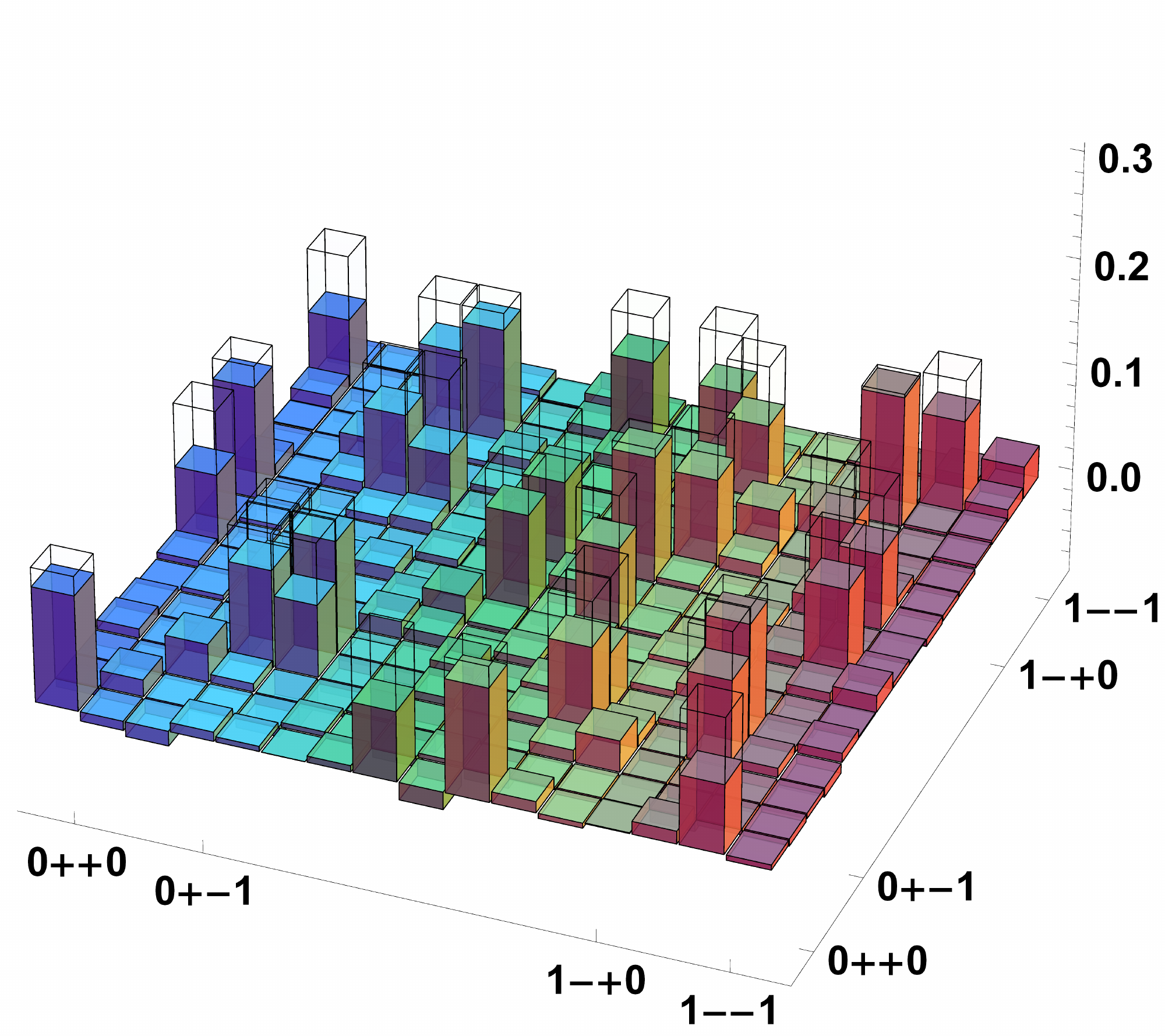}
\caption{Density matrix (real part) of the four-qubit box
cluster state after a half phase ($e^{-i\frac{\pi}{4}Z}$) error occurred on qubit 2 ($F= 0.641 \pm  0.009$ via local unitary operations). 
}
\label{fig:PlotErrHalf2}
\end{figure}

\begin{figure}[H]
\centering
\includegraphics[width=0.33\textwidth]{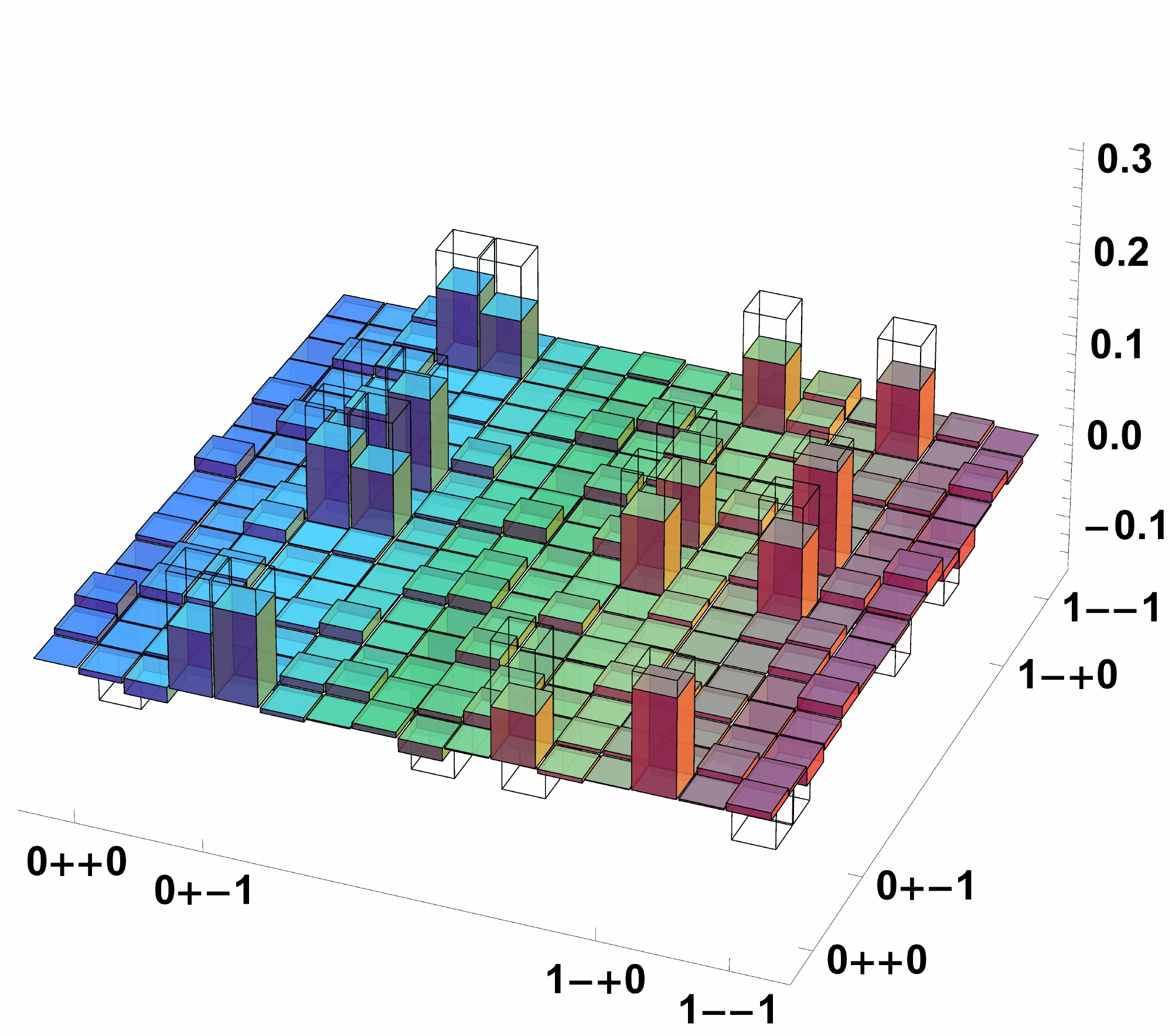}
\caption{Density matrix (imaginary part) of the four-qubit box
cluster state after a half phase ($e^{-i\frac{\pi}{4}Z}$) error occurred on qubit 2 (same fidelity as real part, Figure~\ref{fig:PlotErrHalf2}). }
\end{figure}

%

\section{Appendix D: Decoding Results}
In the tables IV, V, VI, and VII we report fidelities related to decoded qubits per each type of implemented error ($e^{-i\frac{\pi}{2}Z}$ error on qubit 2, $e^{-i\frac{\pi}{2}Z}$ error on qubit 3, $e^{-i\frac{\pi}{4}Z}$ error on qubit 2 and  $e^{-i\frac{\pi}{4}Z}$ error qubit 3) and considering different syndrome outcomes (associated to  applying or not the recovery operation $X$ ). Specifically we report two tables for the $e^{-i\frac{\pi}{2}Z}$ error type, the first related to the outcomes with no need of recovery operation and the second related  to the results after the recovery operation $X$. The same for the $e^{-i\frac{\pi}{4}Z}$ error type.\\
The last column of each table shows fidelities for decoded qubits in the presence of two errors, on both qubit 2 and 3. When this case happens we have to discard the obtained results since the relative syndrome outcomes can be confused with the respectives to one qubit error case.
Further, we show in the Bloch sphere representation the decoded qubit $|M\rangle$ after an  $e^{-i\frac{\pi}{4}Z}$  error acted on qubit 3 (see Fig.~\ref{Figure4-version2}).

\begin{widetext}

\begin{table}[H]
\centering
\begin{tabular}{>{\centering}m{1.5cm}|>{\centering}m{1.5cm}|>{\centering}m{1.5cm}|>{\centering}m{1.5cm}|>{\centering}m{1.5cm}|>{\centering}m{3.2cm}|>{\centering}m{1.5cm}|c|c}
\hline 
\multicolumn{2}{c}{Measured State Lab} &  \multicolumn{2}{c}{Measured State Box} &  \multicolumn{5}{c}{Encoded State Box} \\ 
\hline
Qubit & Basis & Qubit & Basis & \multicolumn{1}{c}{Qubit} & &  Basis & $\theta$ & $\phi$  \\
\hline 
\hline 
$\ket{+}$ &  $\sigma_{x}$ & $\ket{0}$ &  $\sigma_{z}$ &  $\ket{0}$ & & $\sigma_{z}$ & $0\text{\textdegree}$ & $0\text{\textdegree}$  \tabularnewline
\hline 
$\ket{0}$ & $\sigma_{z}$ & $\ket{+}$ & $\sigma_{x}$ & $\ket{+}$   & $\frac{1}{\sqrt{2}}(\ket{0}+\ket{1})$ & $\sigma_{x}$ &  $90\text{\textdegree}$ & $0\text{\textdegree}$ \tabularnewline
\hline 
$\ket{-_{i}}$ & $-\sigma_{y}$  & $\ket{+_{i}}$ & $\sigma_{y}$  & $\ket{-_{i}}$ & $\frac{1}{\sqrt{2}}(\ket{0} - i\ket{1})$ & $-\sigma_{y}$  & $90\text{\textdegree}$ & $-90\text{\textdegree}$ \tabularnewline
\hline 
$\ket{P}$  & $\sigma_{x}+\sigma_{y}$ & $\ket{U}$  & $\sigma_{z}-\sigma_{y}$ & $\ket{T}$ &  $\frac{1}{\sqrt{2}}(\ket{+}+e^{-i\pi/4}\ket{-})$ &  $\sigma_{z}+\sigma_{y}$  & $ 45\text{\textdegree}$ & $90\text{\textdegree}$ \tabularnewline
\hline 
$\ket{M}$  & $\sigma_{x}-\sigma_{y}$ & $\ket{T}$  & $\sigma_{z}+\sigma_{y}$ & $\ket{U}$ & $\frac{1}{\sqrt{2}}(\ket{+}+e^{i\pi/4}\ket{-})$ & $\sigma_{z}-\sigma_{y}$  & $45\text{\textdegree}$  & $-90\text{\textdegree}$ \tabularnewline
\hline 
$\ket{Q}$  & $\sigma_{x}+\sigma_{z}$ & $\ket{Q}$  & $\sigma_{z}+\sigma_{x}$ &  $\ket{Q}$  &    $\frac{1}{\sqrt{2}}(\ket{+_{i}}+e^{i\pi/4}\ket{-_i})$ & $\sigma_{z}+\sigma_{x}$ &  $45\text{\textdegree}$ & $0\text{\textdegree}$  \tabularnewline
\hline 
$\ket{S}$ & $\sigma_{x}-\sigma_{z}$ & $\ket{N}$ & $\sigma_{z}-\sigma_{x}$ & $\ket{N}$  &  $\frac{1}{\sqrt{2}}(\ket{+_{i}}+e^{-i\pi/4}\ket{-_i})$ & $\sigma_{z}-\sigma_{x}$ & $45\text{\textdegree}$ & $180\text{\textdegree}$ \tabularnewline
\hline 
$\ket{T}$ &  $\sigma_{z}+\sigma_{y}$ & $\ket{M}$ & $\sigma_{x}-\sigma_{y}$  & $\ket{P}$ & $\frac{1}{\sqrt{2}}(\ket{0}+e^{+i\pi/4}\ket{1})$  &  $\sigma_{x}+\sigma_{y}$ & $90\text{\textdegree}$ & $45\text{\textdegree}$  \tabularnewline
\hline 
$\ket{U}$ & $\sigma_{z}-\sigma_{y}$  & $\ket{P}$ &  $\sigma_{x}+\sigma_{y}$ & $\ket{M}$  & $\frac{1}{\sqrt{2}}(\ket{0}+e^{-i\pi/4}\ket{1})$ & $\sigma_{x}-\sigma_{y}$  & $90\text{\textdegree}$ & $-45\text{\textdegree}$    \tabularnewline
\hline 
\end{tabular}
\caption{Initial states of qubit 1 reported in different notations: the measured state lab  shows the projection state $\alpha^*|0\rangle + \beta^*|1\rangle$ of qubit 1 of $\ket{\psi_{lab}}$ (see main paper) and relative basis; the measured state box shows the projection state $\alpha^*|0\rangle + \beta^*|1\rangle$ of qubit 1 of $\ket{\psi_{box}}$ and relative basis; and the encoded state box notation shows the encoded qubit 1 in the box cluster notation. In this case we report per input state the explicit qubit, the relative basis and the spherical coordinates. \label{tab:Encoded-States}}
\end{table}

\begin{table}[H]
  \centering
   \begin{tabular}{m{2.4cm}cccc}
 \hline
    \textbf{Encoded State}  & \textbf{NO Error} &\textbf{Error}   $Z_3$  &  \textbf{Error} $Z_2$  & \textbf{Error} $Z_2Z_3$\\
 \hline
$\ket{0}$   & F=0.967 $\pm$ 0.011 & F=0.874 $\pm$ 0.022 & F=0.904 $\pm$ 0.021 & F=0.886 $\pm$ 0.022 \\
$\ket{+}$    & F=0.944 $\pm$ 0.017 & F=0.945 $\pm$ 0.016 & F=0.943 $\pm$ 0.015 & F=0.969$\pm$ 0.009 \\
$\ket{-_{i}}$     & F=0.956 $\pm$ 0.012 & F=0.925 $\pm$ 0.019 & F=0.912 $\pm$ 0.019 & F=0.905 $\pm$ 0.019 \\
$\ket{T}$  & F=0.820 $\pm$ 0.030 & F=0.965 $\pm$ 0.021 & F=0.971 $\pm$ 0.014 & F=0.907 $\pm$ 0.028 \\
$\ket{U}$    & F=0.863 $\pm$ 0.025 & F=0.957 $\pm$ 0.013 & F=0.927 $\pm$ 0.025 & F=0.885 $\pm$ 0.028 \\
$\ket{Q}$  & F=0.938 $\pm$ 0.026 & F=0. 898$\pm$ 0.033 & F=0.915 $\pm$ 0.026 & F=0.976 $\pm$ 0.009 \\
$\ket{N}$  & F=0.967 $\pm$ 0.020 & F=0.940 $\pm$ 0.024 & F=0.915 $\pm$ 0.026 & F=0.922 $\pm$ 0.024 \\
$\ket{P}$   & F=0.895 $\pm$ 0.033 & F=0.902 $\pm$ 0.026 & F=0.810 $\pm$ 0.036 & F=0.910 $\pm$ 0.028 \\
$\ket{M}$  & F=0.918 $\pm$ 0.026 & F=0.965 $\pm$ 0.016 & F=0.947 $\pm$ 0.023 & F=0.970 $\pm$ 0.018 \\
 \hline
    \end{tabular} 
\caption{Fidelities of different encoded and decoded states after an $e^{-i\frac{\pi}{2}Z}$ error was implemented, but no recovery operation was necessary.\label{TableFid1}}
\end{table}

\begin{table}[H]
  \centering
   \begin{tabular}{m{2.4cm}cccc}
 \hline
  \textbf{Encoded State}  & \textbf{NO Error} &\textbf{Error}   $Z_3$  &  \textbf{Error} $Z_2$  & \textbf{Error} $Z_2Z_3$\\
 \hline
  $\ket{0}$    & F=0.767 $\pm$ 0.032 & F=0.716 $\pm$ 0.035 & F=0.793 $\pm$ 0.030 & F=0.784 $\pm$ 0.028 \\
 $\ket{+}$     & F=0.948 $\pm$ 0.016 & F=0.945 $\pm$ 0.018 & F=0.944 $\pm$ 0.014 & F=0.943 $\pm$ 0.015 \\
$\ket{-_{i}}$     & F=0.684 $\pm$ 0.036 & F=0.674 $\pm$ 0.038 & F=0.710 $\pm$ 0.036 & F=0.658 $\pm$ 0.034 \\
$\ket{T}$  & F=0.686 $\pm$ 0.033 & F=0.809 $\pm$ 0.031 & F=0.792 $\pm$ 0.032 & F=0.700 $\pm$ 0.028 \\
$\ket{U}$    & F=0.681 $\pm$ 0.037 & F=0.771 $\pm$ 0.032 & F=0.662 $\pm$ 0.034 & F=0.740 $\pm$ 0.038 \\
$\ket{Q}$  & F=0.741 $\pm$ 0.044 & F=0.776 $\pm$ 0.034 & F=0.833 $\pm$ 0.031 & F=0.813 $\pm$ 0.029 \\
$\ket{N}$   & F=0.793 $\pm$ 0.047 & F=0.875 $\pm$ 0.029 & F=0.843 $\pm$ 0.029 & F=0.962 $\pm$ 0.027 \\
$\ket{P}$   & F=0.750 $\pm$ 0.029 & F=0.811 $\pm$ 0.031 & F=0.790 $\pm$ 0.028 & F=0.862 $\pm$ 0.024 \\
$\ket{M}$   & F=0.800 $\pm$ 0.035 & F=0.801 $\pm$ 0.038 & F=0.823 $\pm$ 0.029 & F=0.895 $\pm$ 0.030 \\
 \hline
    \end{tabular}
\caption{Fidelities of different encoded and decoded states after an $e^{-i\frac{\pi}{2}Z}$ error was implemented and a recovery operation performed.\label{TableFid2}}
 \end{table}

\begin{table}[H]
  \centering
  \begin{tabular}{m{2.4cm}cccc}
     \hline
  \textbf{Encoded State} & \multicolumn{2}{c}{$e^{-i\frac{\pi}{4}Z_3}$ \textbf{Error}} & \multicolumn{2}{c}{$e^{-i\frac{\pi}{4}Z_2}$ \textbf{Error}}\\ 
     \hline
            & \textbf{NO Error *} &\textbf{Error}   $Z_3$  &  \textbf{ NO Error} & \textbf{Error *} $Z_2$\\
\hline
 $\ket{0}$      & F=0.896 $\pm$ 0.027 & F=0.865 $\pm$ 0.035 & F=0.909 $\pm$ 0.021 & F=0.927 $\pm$ 0.015 \\
 $\ket{+}$      & F=0.861 $\pm$ 0.034 & F=0.913 $\pm$ 0.022 & F=0.936 $\pm$ 0.021 & F=0.986 $\pm$ 0.007 \\
 $\ket{-_{i}}$  & F=0.908 $\pm$ 0.024 & F=0.865 $\pm$ 0.025 & F=0.945 $\pm$ 0.017 & F=0.914 $\pm$ 0.022 \\
$\ket{T}$   		& F=0.850 $\pm$ 0.042 & F=0.880 $\pm$ 0.040 & F=0.823 $\pm$ 0.036 & F=0.842 $\pm$ 0.030 \\
$\ket{U}$   		& F=0.951 $\pm$ 0.034 & F=0.951 $\pm$ 0.013 & F=0.930 $\pm$ 0.028 & F=0.903 $\pm$ 0.028 \\
$\ket{Q}$ 		 	& F=0.849 $\pm$ 0.037 & F=0.854 $\pm$ 0.035 & F=0.905 $\pm$ 0.031 & F=0.937 $\pm$ 0.028 \\
$\ket{N}$ 		 	& F=0.963 $\pm$ 0.023 & F=0.942 $\pm$ 0.024 & F=0.894 $\pm$ 0.027 & F=0.923 $\pm$ 0.025 \\
$\ket{P}$ 		 	& F=0.871 $\pm$ 0.030 & F=0.964 $\pm$ 0.021 & F=0.972 $\pm$ 0.011 & F=0.972 $\pm$ 0.015 \\
$\ket{M}$ 		 	& F=0.956 $\pm$ 0.022 & F=0.984 $\pm$ 0.016 & F=0.990 $\pm$ 0.009 & F=0.929 $\pm$ 0.025 \\
     \hline
    \end{tabular}%
 \caption{Fidelities of different encoded and decoded states after an $e^{-i\frac{\pi}{4}Z}$ error was implemented, but no recovery operation performed.\label{TableFid3} }
\end{table}

\begin{table}[H]
\centering
    \begin{tabular}{m{2.4cm}cccc}
        \hline
  \multicolumn{1}{l}{\textbf{Encoded State}} &  \multicolumn{2}{c}{$e^{-i\frac{\pi}{4}Z_3}$ \textbf{Error}} & \multicolumn{2}{c}{$e^{-i\frac{\pi}{4}Z_2}$ \textbf{Error}}\\ 
  \hline
            & \textbf{NO Error *} &\textbf{Error}   $Z_3$  &  \textbf{ NO Error} & \textbf{Error *} $Z_2$\\
\hline
$\ket{0}$      		&F=0.702 $\pm$ 0.038 &F=0.843 $\pm$ 0.280 &F=0.690 $\pm$ 0.036 &F=0.721 $\pm$ 0.029 \\
$\ket{+}$       	&F=0.948 $\pm$ 0.015 &F=0.981 $\pm$ 0.008 &F=0.896 $\pm$ 0.024 &F=0.837$\pm$ 0.023 \\
$\ket{-_{i}}$    &F=0.629. $\pm$ 0.039 &F=0.814 $\pm$ 0.035 &F=0.697 $\pm$ 0.040 &F=0.644 $\pm$ 0.030 \\
$\ket{T}$    	  	&F=0.658 $\pm$ 0.033 &F=0.691 $\pm$ 0.038 &F=0.699 $\pm$ 0.037 &F=0.656 $\pm$ 0.033 \\
$\ket{U}$   			&F=0.677 $\pm$ 0.032 &F=0.814 $\pm$ 0.037 &F=0.813 $\pm$ 0.039 &F=0.776 $\pm$ 0.034 \\
$\ket{Q}$   		  &F=0.832 $\pm$ 0.032 &F=0.818 $\pm$ 0.031 &F=0.842 $\pm$ 0.0345 &F=0.734 $\pm$ 0.033 \\
$\ket{N}$   		  &F=0.932 $\pm$ 0.028 &F=0.857 $\pm$ 0.038 &F=0.794 $\pm$ 0.041 &F=0.894 $\pm$ 0.031 \\
$\ket{P}$  	   		&F=0.749 $\pm$ 0.029 &F=0.919 $\pm$ 0.028 &F=0.799 $\pm$ 0.031 &F=0.706 $\pm$ 0.031 \\
$\ket{M}$  		 		&F=0.775 $\pm$ 0.034 &F=0.941 $\pm$ 0.025 &F=0.906$\pm$ 0.028 &F=0.823 $\pm$ 0.032\\
         \hline
    \end{tabular}
 \caption{Fidelities of different encoded and decoded states after an $e^{-i\frac{\pi}{4}Z}$ error was implemented and a recovery operation performed.\label{TableFid4}}
\end{table}


\begin{figure}[H]
\centering
\includegraphics[width=0.90\textwidth]{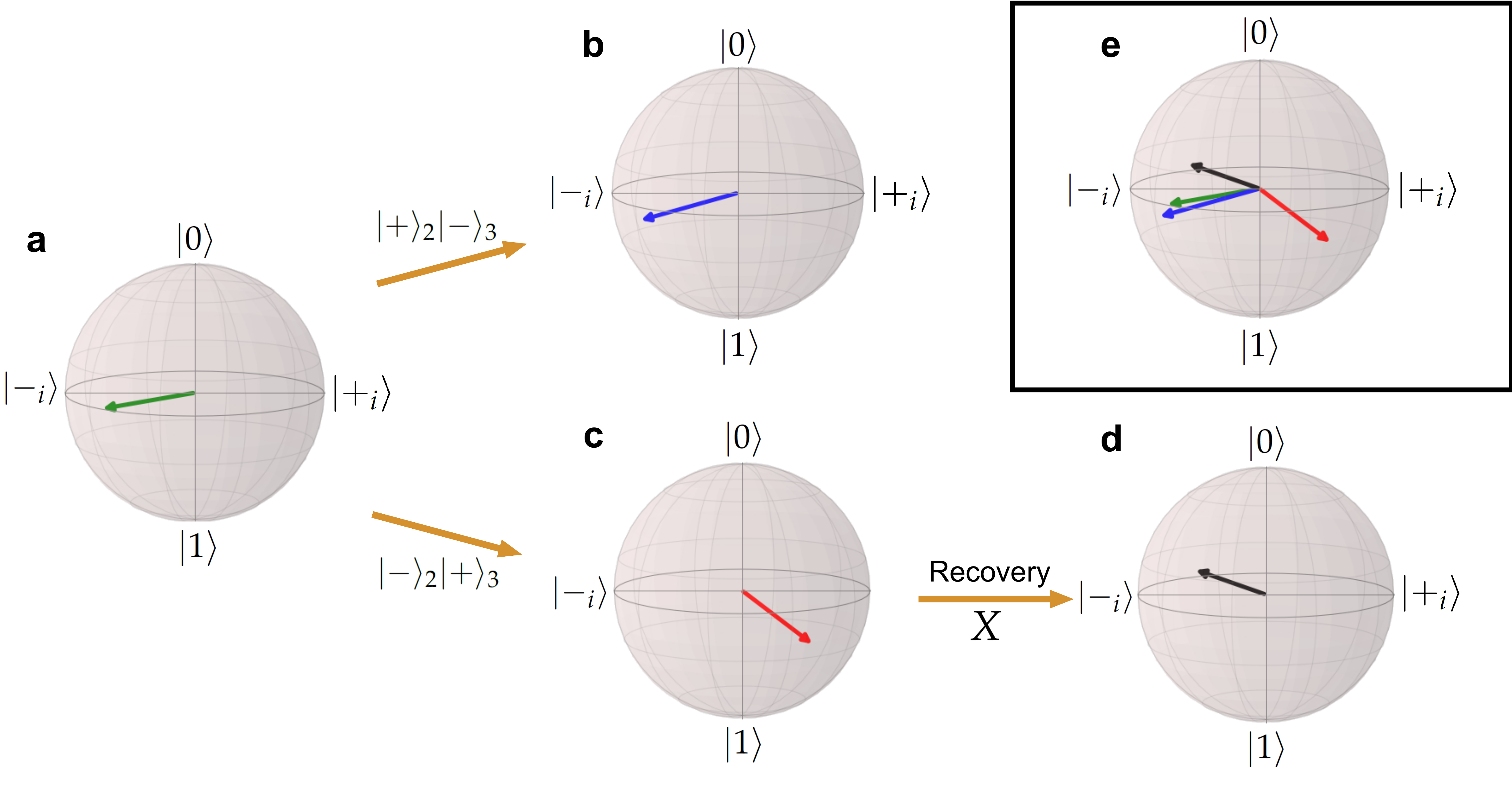}
\caption{\label{Figure4-version2} Experimental decoding results in Bloch sphere Representation for a $e^{-i\frac{\pi}{4}Z}$  error acted on qubit 3.
\textbf{a,} Encoded/ideal state  $\ket{M}$, an eigenstate of the operator $(X-Y)$.
\textbf{b,} Decoded qubit reading the outcomes $|-+\rangle$ of qubit 2 and 3, $F=0.984\pm0.016$.
\textbf{c,d,} Decoded qubit reading the outcomes $|+-\rangle$ before and after applying the recovery operation $X$, $F=0.941\pm0.025$.
\textbf{e,} Overview of the encoded and decoded qubits.
}
\end{figure}
\end{widetext}

\begin{thebibliography}{26}
\expandafter\ifx\csname natexlab\endcsname\relax\def\natexlab#1{#1}\fi
\expandafter\ifx\csname bibnamefont\endcsname\relax
  \def\bibnamefont#1{#1}\fi
\expandafter\ifx\csname bibfnamefont\endcsname\relax
  \def\bibfnamefont#1{#1}\fi
\expandafter\ifx\csname citenamefont\endcsname\relax
  \def\citenamefont#1{#1}\fi
\expandafter\ifx\csname url\endcsname\relax
  \def\url#1{\texttt{#1}}\fi
\expandafter\ifx\csname urlprefix\endcsname\relax\def\urlprefix{URL }\fi
\providecommand{\bibinfo}[2]{#2}
\providecommand{\eprint}[2][]{\url{#2}}

\bibitem[{\citenamefont{Raussendorf and Briegel}(2001)}]{Raussendorf2001}
\bibinfo{author}{\bibfnamefont{R.}~\bibnamefont{Raussendorf}} \bibnamefont{and}
  \bibinfo{author}{\bibfnamefont{H.}~\bibnamefont{Briegel}},
  \bibinfo{journal}{Phys. Rev. Lett.} \textbf{\bibinfo{volume}{86}},
  \bibinfo{pages}{5188} (\bibinfo{year}{2001}).

\bibitem[{\citenamefont{Zhang and Braunstein}(2006)}]{Zhang2006b}
\bibinfo{author}{\bibfnamefont{J.}~\bibnamefont{Zhang}} \bibnamefont{and}
  \bibinfo{author}{\bibfnamefont{S.~L.} \bibnamefont{Braunstein}},
  \bibinfo{journal}{Physical Review A} \textbf{\bibinfo{volume}{73}},
  \bibinfo{pages}{032318} (\bibinfo{year}{2006}).

\bibitem[{\citenamefont{Menicucci et~al.}(2006)\citenamefont{Menicucci, van
  Loock, Gu, Weedbrook, Ralph, and Nielsen}}]{Menicucci2006}
\bibinfo{author}{\bibfnamefont{N.~C.} \bibnamefont{Menicucci}},
  \bibinfo{author}{\bibfnamefont{P.}~\bibnamefont{van Loock}},
  \bibinfo{author}{\bibfnamefont{M.}~\bibnamefont{Gu}},
  \bibinfo{author}{\bibfnamefont{C.}~\bibnamefont{Weedbrook}},
  \bibinfo{author}{\bibfnamefont{T.~C.} \bibnamefont{Ralph}}, \bibnamefont{and}
  \bibinfo{author}{\bibfnamefont{M.~A.} \bibnamefont{Nielsen}},
  \bibinfo{journal}{Physical review letters} \textbf{\bibinfo{volume}{97}},
  \bibinfo{pages}{110501} (\bibinfo{year}{2006}).

\bibitem[{\citenamefont{Briegel et~al.}(2009)\citenamefont{Briegel, Browne,
  D{\"u}r, Raussendorf, and Van~den Nest}}]{Briegel2009}
\bibinfo{author}{\bibfnamefont{H.-J.} \bibnamefont{Briegel}},
  \bibinfo{author}{\bibfnamefont{D.~E.} \bibnamefont{Browne}},
  \bibinfo{author}{\bibfnamefont{W.}~\bibnamefont{D{\"u}r}},
  \bibinfo{author}{\bibfnamefont{R.}~\bibnamefont{Raussendorf}},
  \bibnamefont{and} \bibinfo{author}{\bibfnamefont{M.}~\bibnamefont{Van~den
  Nest}}, \bibinfo{journal}{Nature Phys.} \textbf{\bibinfo{volume}{5}},
  \bibinfo{pages}{19} (\bibinfo{year}{2009}), ISSN \bibinfo{issn}{1745-2473}.

\bibitem[{\citenamefont{Briegel and Raussendorf}(2001)}]{Briegel2001}
\bibinfo{author}{\bibfnamefont{H.~J.} \bibnamefont{Briegel}} \bibnamefont{and}
  \bibinfo{author}{\bibfnamefont{R.}~\bibnamefont{Raussendorf}},
  \bibinfo{journal}{Phys.\ Rev.\ Lett.} \textbf{\bibinfo{volume}{86}},
  \bibinfo{pages}{910} (\bibinfo{year}{2001}).

\bibitem[{\citenamefont{Raussendorf et~al.}(2003)\citenamefont{Raussendorf,
  Browne, and Briegel}}]{Raussendorf2003}
\bibinfo{author}{\bibfnamefont{R.}~\bibnamefont{Raussendorf}},
  \bibinfo{author}{\bibfnamefont{D.~E.} \bibnamefont{Browne}},
  \bibnamefont{and} \bibinfo{author}{\bibfnamefont{H.~J.}
  \bibnamefont{Briegel}}, \bibinfo{journal}{Phys. Rev. A}
  \textbf{\bibinfo{volume}{68}}, \bibinfo{pages}{022312}
  (\bibinfo{year}{2003}).

\bibitem[{\citenamefont{Zwerger et~al.}(2012)\citenamefont{Zwerger, D{\"u}r,
  and Briegel}}]{Zwerger2012}
\bibinfo{author}{\bibfnamefont{M.}~\bibnamefont{Zwerger}},
  \bibinfo{author}{\bibfnamefont{W.}~\bibnamefont{D{\"u}r}}, \bibnamefont{and}
  \bibinfo{author}{\bibfnamefont{H.}~\bibnamefont{Briegel}},
  \bibinfo{journal}{Phys. Rev. A} \textbf{\bibinfo{volume}{85}},
  \bibinfo{pages}{062326} (\bibinfo{year}{2012}).

\bibitem[{\citenamefont{Gottesman}(1997)}]{Gottesman1997}
\bibinfo{author}{\bibfnamefont{D.}~\bibnamefont{Gottesman}}, Ph.D. thesis,
  \bibinfo{school}{Caltech} (\bibinfo{year}{1997}).

\bibitem[{\citenamefont{Zwerger
  et~al.}(2013{\natexlab{a}})\citenamefont{Zwerger, Briegel, and
  D\"ur}}]{Zwerger2013}
\bibinfo{author}{\bibfnamefont{M.}~\bibnamefont{Zwerger}},
  \bibinfo{author}{\bibfnamefont{H.~J.} \bibnamefont{Briegel}},
  \bibnamefont{and} \bibinfo{author}{\bibfnamefont{W.}~\bibnamefont{D\"ur}},
  \bibinfo{journal}{Phys. Rev. Lett.} \textbf{\bibinfo{volume}{110}},
  \bibinfo{pages}{260503} (\bibinfo{year}{2013}{\natexlab{a}}).

\bibitem[{\citenamefont{Zwerger
  et~al.}(2013{\natexlab{b}})\citenamefont{Zwerger, Briegel, and
  D\"ur}}]{Zwerger2013a}
\bibinfo{author}{\bibfnamefont{M.}~\bibnamefont{Zwerger}},
  \bibinfo{author}{\bibfnamefont{H.~J.} \bibnamefont{Briegel}},
  \bibnamefont{and} \bibinfo{author}{\bibfnamefont{W.}~\bibnamefont{D\"ur}},
  \bibinfo{journal}{arXiv: 1308.4561}  (\bibinfo{year}{2013}{\natexlab{b}}).

\bibitem[{\citenamefont{Walther et~al.}(2005)\citenamefont{Walther, Resch,
  Rudolph, Schenck, Weinfurter, Vedral, Aspelmeyer, and
  Zeilinger}}]{Walther2005a}
\bibinfo{author}{\bibfnamefont{P.}~\bibnamefont{Walther}},
  \bibinfo{author}{\bibfnamefont{K.}~\bibnamefont{Resch}},
  \bibinfo{author}{\bibfnamefont{T.}~\bibnamefont{Rudolph}},
  \bibinfo{author}{\bibfnamefont{E.}~\bibnamefont{Schenck}},
  \bibinfo{author}{\bibfnamefont{H.}~\bibnamefont{Weinfurter}},
  \bibinfo{author}{\bibfnamefont{V.}~\bibnamefont{Vedral}},
  \bibinfo{author}{\bibfnamefont{M.}~\bibnamefont{Aspelmeyer}},
  \bibnamefont{and}
  \bibinfo{author}{\bibfnamefont{A.}~\bibnamefont{Zeilinger}},
  \bibinfo{journal}{Nature} \textbf{\bibinfo{volume}{434}},
  \bibinfo{pages}{169} (\bibinfo{year}{2005}).

\bibitem[{\citenamefont{Chen et~al.}(2007)\citenamefont{Chen, Li, Zhang, Chen,
  Goebel, Chen, Mair, and Pan}}]{Chen2007}
\bibinfo{author}{\bibfnamefont{K.}~\bibnamefont{Chen}},
  \bibinfo{author}{\bibfnamefont{C.-M.} \bibnamefont{Li}},
  \bibinfo{author}{\bibfnamefont{Q.}~\bibnamefont{Zhang}},
  \bibinfo{author}{\bibfnamefont{Y.-A.} \bibnamefont{Chen}},
  \bibinfo{author}{\bibfnamefont{A.}~\bibnamefont{Goebel}},
  \bibinfo{author}{\bibfnamefont{S.}~\bibnamefont{Chen}},
  \bibinfo{author}{\bibfnamefont{A.}~\bibnamefont{Mair}}, \bibnamefont{and}
  \bibinfo{author}{\bibfnamefont{J.-W.} \bibnamefont{Pan}},
  \bibinfo{journal}{Phys. Rev. Lett.} \textbf{\bibinfo{volume}{99}},
  \bibinfo{eid}{120503} (pages~\bibinfo{numpages}{4}) (\bibinfo{year}{2007}).

\bibitem[{\citenamefont{Lu et~al.}(2007)\citenamefont{Lu, Zhou, G\"{u}hne, Gao,
  Zhang, Yuan, Goebel, Yang, and Pan}}]{Lu2007a}
\bibinfo{author}{\bibfnamefont{C.}~\bibnamefont{Lu}},
  \bibinfo{author}{\bibfnamefont{X.}~\bibnamefont{Zhou}},
  \bibinfo{author}{\bibfnamefont{O.}~\bibnamefont{G\"{u}hne}},
  \bibinfo{author}{\bibfnamefont{W.}~\bibnamefont{Gao}},
  \bibinfo{author}{\bibfnamefont{J.}~\bibnamefont{Zhang}},
  \bibinfo{author}{\bibfnamefont{Z.}~\bibnamefont{Yuan}},
  \bibinfo{author}{\bibfnamefont{A.}~\bibnamefont{Goebel}},
  \bibinfo{author}{\bibfnamefont{T.}~\bibnamefont{Yang}}, \bibnamefont{and}
  \bibinfo{author}{\bibfnamefont{J.}~\bibnamefont{Pan}},
  \bibinfo{journal}{Nature Phys.} \textbf{\bibinfo{volume}{3}},
  \bibinfo{pages}{91} (\bibinfo{year}{2007}).

\bibitem[{\citenamefont{Tokunaga et~al.}(2008)\citenamefont{Tokunaga,
  Kuwashiro, Yamamoto, Koashi, and Imoto}}]{Tokunaga2008}
\bibinfo{author}{\bibfnamefont{Y.}~\bibnamefont{Tokunaga}},
  \bibinfo{author}{\bibfnamefont{S.}~\bibnamefont{Kuwashiro}},
  \bibinfo{author}{\bibfnamefont{T.}~\bibnamefont{Yamamoto}},
  \bibinfo{author}{\bibfnamefont{M.}~\bibnamefont{Koashi}}, \bibnamefont{and}
  \bibinfo{author}{\bibfnamefont{N.}~\bibnamefont{Imoto}},
  \bibinfo{journal}{Phys. Rev. Lett.} \textbf{\bibinfo{volume}{100}},
  \bibinfo{pages}{210501} (\bibinfo{year}{2008}), ISSN
  \bibinfo{issn}{1079-7114}.

\bibitem[{\citenamefont{Vallone et~al.}(2008)\citenamefont{Vallone, Pomarico,
  Martini, and Mataloni}}]{Vallone2008a}
\bibinfo{author}{\bibfnamefont{G.}~\bibnamefont{Vallone}},
  \bibinfo{author}{\bibfnamefont{E.}~\bibnamefont{Pomarico}},
  \bibinfo{author}{\bibfnamefont{F.~D.} \bibnamefont{Martini}},
  \bibnamefont{and} \bibinfo{author}{\bibfnamefont{P.}~\bibnamefont{Mataloni}},
  \bibinfo{journal}{Phys. Rev. A} \textbf{\bibinfo{volume}{78}}
  (\bibinfo{year}{2008}).

\bibitem[{\citenamefont{Vallone et~al.}(2010)\citenamefont{Vallone, Donati,
  Bruno, Chiuri, and Mataloni}}]{Vallone2010}
\bibinfo{author}{\bibfnamefont{G.}~\bibnamefont{Vallone}},
  \bibinfo{author}{\bibfnamefont{G.}~\bibnamefont{Donati}},
  \bibinfo{author}{\bibfnamefont{N.}~\bibnamefont{Bruno}},
  \bibinfo{author}{\bibfnamefont{A.}~\bibnamefont{Chiuri}}, \bibnamefont{and}
  \bibinfo{author}{\bibfnamefont{P.}~\bibnamefont{Mataloni}},
  \bibinfo{journal}{Phys. Rev. A} \textbf{\bibinfo{volume}{81}},
  \bibinfo{pages}{50302} (\bibinfo{year}{2010}), ISSN
  \bibinfo{issn}{1094-1622}.

\bibitem[{\citenamefont{Ukai et~al.}(2011{\natexlab{a}})\citenamefont{Ukai,
  Iwata, Shimokawa, Armstrong, Politi, Yoshikawa, van Loock, and
  Furusawa}}]{Ukai2011}
\bibinfo{author}{\bibfnamefont{R.}~\bibnamefont{Ukai}},
  \bibinfo{author}{\bibfnamefont{N.}~\bibnamefont{Iwata}},
  \bibinfo{author}{\bibfnamefont{Y.}~\bibnamefont{Shimokawa}},
  \bibinfo{author}{\bibfnamefont{S.~C.} \bibnamefont{Armstrong}},
  \bibinfo{author}{\bibfnamefont{A.}~\bibnamefont{Politi}},
  \bibinfo{author}{\bibfnamefont{J.-i.} \bibnamefont{Yoshikawa}},
  \bibinfo{author}{\bibfnamefont{P.}~\bibnamefont{van Loock}},
  \bibnamefont{and} \bibinfo{author}{\bibfnamefont{A.}~\bibnamefont{Furusawa}},
  \bibinfo{journal}{Physical review letters} \textbf{\bibinfo{volume}{106}},
  \bibinfo{pages}{240504} (\bibinfo{year}{2011}{\natexlab{a}}).

\bibitem[{\citenamefont{Ukai et~al.}(2011{\natexlab{b}})\citenamefont{Ukai,
  Yokoyama, Yoshikawa, van Loock, and Furusawa}}]{Ukai2011a}
\bibinfo{author}{\bibfnamefont{R.}~\bibnamefont{Ukai}},
  \bibinfo{author}{\bibfnamefont{S.}~\bibnamefont{Yokoyama}},
  \bibinfo{author}{\bibfnamefont{J.-i.} \bibnamefont{Yoshikawa}},
  \bibinfo{author}{\bibfnamefont{P.}~\bibnamefont{van Loock}},
  \bibnamefont{and} \bibinfo{author}{\bibfnamefont{A.}~\bibnamefont{Furusawa}},
  \bibinfo{journal}{Physical Review Letters} \textbf{\bibinfo{volume}{107}},
  \bibinfo{pages}{250501} (\bibinfo{year}{2011}{\natexlab{b}}).

\bibitem[{\citenamefont{Yao et~al.}(2012{\natexlab{a}})\citenamefont{Yao, Wang,
  Xu, Lu, Pan, Bao, Peng, Lu, Chen, and Pan}}]{Yao2012a}
\bibinfo{author}{\bibfnamefont{X.-C.} \bibnamefont{Yao}},
  \bibinfo{author}{\bibfnamefont{T.-X.} \bibnamefont{Wang}},
  \bibinfo{author}{\bibfnamefont{P.}~\bibnamefont{Xu}},
  \bibinfo{author}{\bibfnamefont{H.}~\bibnamefont{Lu}},
  \bibinfo{author}{\bibfnamefont{G.-S.} \bibnamefont{Pan}},
  \bibinfo{author}{\bibfnamefont{X.-H.} \bibnamefont{Bao}},
  \bibinfo{author}{\bibfnamefont{C.-Z.} \bibnamefont{Peng}},
  \bibinfo{author}{\bibfnamefont{C.-Y.} \bibnamefont{Lu}},
  \bibinfo{author}{\bibfnamefont{Y.-A.} \bibnamefont{Chen}}, \bibnamefont{and}
  \bibinfo{author}{\bibfnamefont{J.-W.} \bibnamefont{Pan}},
  \bibinfo{journal}{Nature Photonics} \textbf{\bibinfo{volume}{6}},
  \bibinfo{pages}{225} (\bibinfo{year}{2012}{\natexlab{a}}).

\bibitem[{\citenamefont{Su et~al.}(2012)\citenamefont{Su, Zhao, Hao, Jia, Xie,
  and Peng}}]{Su2012}
\bibinfo{author}{\bibfnamefont{X.}~\bibnamefont{Su}},
  \bibinfo{author}{\bibfnamefont{Y.}~\bibnamefont{Zhao}},
  \bibinfo{author}{\bibfnamefont{S.}~\bibnamefont{Hao}},
  \bibinfo{author}{\bibfnamefont{X.}~\bibnamefont{Jia}},
  \bibinfo{author}{\bibfnamefont{C.}~\bibnamefont{Xie}}, \bibnamefont{and}
  \bibinfo{author}{\bibfnamefont{K.}~\bibnamefont{Peng}},
  \bibinfo{journal}{Optics letters} \textbf{\bibinfo{volume}{37}},
  \bibinfo{pages}{5178} (\bibinfo{year}{2012}).

\bibitem[{\citenamefont{Yokoyama et~al.}(2013)\citenamefont{Yokoyama, Ukai,
  Armstrong, Sornphiphatphong, Kaji, Suzuki, Yoshikawa, Yonezawa, Menicucci,
  and Furusawa}}]{Yokoyama2013}
\bibinfo{author}{\bibfnamefont{S.}~\bibnamefont{Yokoyama}},
  \bibinfo{author}{\bibfnamefont{R.}~\bibnamefont{Ukai}},
  \bibinfo{author}{\bibfnamefont{S.~C.} \bibnamefont{Armstrong}},
  \bibinfo{author}{\bibfnamefont{C.}~\bibnamefont{Sornphiphatphong}},
  \bibinfo{author}{\bibfnamefont{T.}~\bibnamefont{Kaji}},
  \bibinfo{author}{\bibfnamefont{S.}~\bibnamefont{Suzuki}},
  \bibinfo{author}{\bibfnamefont{J.-i.} \bibnamefont{Yoshikawa}},
  \bibinfo{author}{\bibfnamefont{H.}~\bibnamefont{Yonezawa}},
  \bibinfo{author}{\bibfnamefont{N.~C.} \bibnamefont{Menicucci}},
  \bibnamefont{and} \bibinfo{author}{\bibfnamefont{A.}~\bibnamefont{Furusawa}},
  \bibinfo{journal}{arXiv preprint arXiv:1306.3366}  (\bibinfo{year}{2013}).

\bibitem[{\citenamefont{Prevedel et~al.}(2007)\citenamefont{Prevedel, Walther,
  Tiefenbacher, B\"{o}hi, Kaltenbaek, Jennewein, and
  Zeilinger}}]{Prevedel2007a}
\bibinfo{author}{\bibfnamefont{R.}~\bibnamefont{Prevedel}},
  \bibinfo{author}{\bibfnamefont{P.}~\bibnamefont{Walther}},
  \bibinfo{author}{\bibfnamefont{F.}~\bibnamefont{Tiefenbacher}},
  \bibinfo{author}{\bibfnamefont{P.}~\bibnamefont{B\"{o}hi}},
  \bibinfo{author}{\bibfnamefont{R.}~\bibnamefont{Kaltenbaek}},
  \bibinfo{author}{\bibfnamefont{T.}~\bibnamefont{Jennewein}},
  \bibnamefont{and}
  \bibinfo{author}{\bibfnamefont{A.}~\bibnamefont{Zeilinger}},
  \bibinfo{journal}{Nature} \textbf{\bibinfo{volume}{445}}, \bibinfo{pages}{65}
  (\bibinfo{year}{2007}).

\bibitem[{\citenamefont{Tame et~al.}(2007)\citenamefont{Tame, Prevedel,
  Paternostro, B\"{o}hi, Kim, and Zeilinger}}]{Tame2007}
\bibinfo{author}{\bibfnamefont{M.~S.} \bibnamefont{Tame}},
  \bibinfo{author}{\bibfnamefont{R.}~\bibnamefont{Prevedel}},
  \bibinfo{author}{\bibfnamefont{M.}~\bibnamefont{Paternostro}},
  \bibinfo{author}{\bibfnamefont{P.}~\bibnamefont{B\"{o}hi}},
  \bibinfo{author}{\bibfnamefont{M.~S.} \bibnamefont{Kim}}, \bibnamefont{and}
  \bibinfo{author}{\bibfnamefont{A.}~\bibnamefont{Zeilinger}},
  \bibinfo{journal}{Phys. Rev. Lett.} \textbf{\bibinfo{volume}{98}},
  \bibinfo{eid}{140501} (pages~\bibinfo{numpages}{4}) (\bibinfo{year}{2007}).

\bibitem[{\citenamefont{Zhao et~al.}(2004)\citenamefont{Zhao, Chen, Zhang,
  Yang, Briegel, and Pan}}]{Zhao2004}
\bibinfo{author}{\bibfnamefont{Z.}~\bibnamefont{Zhao}},
  \bibinfo{author}{\bibfnamefont{Y.-A.} \bibnamefont{Chen}},
  \bibinfo{author}{\bibfnamefont{A.-N.} \bibnamefont{Zhang}},
  \bibinfo{author}{\bibfnamefont{T.}~\bibnamefont{Yang}},
  \bibinfo{author}{\bibfnamefont{H.~J.} \bibnamefont{Briegel}},
  \bibnamefont{and} \bibinfo{author}{\bibfnamefont{J.-W.} \bibnamefont{Pan}},
  \bibinfo{journal}{Nature} \textbf{\bibinfo{volume}{430}}, \bibinfo{pages}{54}
  (\bibinfo{year}{2004}).

\bibitem[{\citenamefont{Yao et~al.}(2012{\natexlab{b}})\citenamefont{Yao, Wang,
  Chen, Gao, Fowler, Raussendorf, Chen, Liu, Lu, Deng et~al.}}]{Yao2012}
\bibinfo{author}{\bibfnamefont{X.-C.} \bibnamefont{Yao}},
  \bibinfo{author}{\bibfnamefont{T.-X.} \bibnamefont{Wang}},
  \bibinfo{author}{\bibfnamefont{H.-Z.} \bibnamefont{Chen}},
  \bibinfo{author}{\bibfnamefont{W.-B.} \bibnamefont{Gao}},
  \bibinfo{author}{\bibfnamefont{A.~G.} \bibnamefont{Fowler}},
  \bibinfo{author}{\bibfnamefont{R.}~\bibnamefont{Raussendorf}},
  \bibinfo{author}{\bibfnamefont{Z.-B.} \bibnamefont{Chen}},
  \bibinfo{author}{\bibfnamefont{N.-L.} \bibnamefont{Liu}},
  \bibinfo{author}{\bibfnamefont{C.-Y.} \bibnamefont{Lu}},
  \bibinfo{author}{\bibfnamefont{Y.-J.} \bibnamefont{Deng}},
  \bibnamefont{et~al.}, \bibinfo{journal}{Nature}
  \textbf{\bibinfo{volume}{482}}, \bibinfo{pages}{489}
  (\bibinfo{year}{2012}{\natexlab{b}}).

\bibitem[{\citenamefont{Lanyon et~al.}(2013)\citenamefont{Lanyon, Jurcevic,
  Zwerger, Hempel, Martinez, D{\"u}r, Briegel, Blatt, and Roos}}]{Lanyon2013}
\bibinfo{author}{\bibfnamefont{B.}~\bibnamefont{Lanyon}},
  \bibinfo{author}{\bibfnamefont{P.}~\bibnamefont{Jurcevic}},
  \bibinfo{author}{\bibfnamefont{M.}~\bibnamefont{Zwerger}},
  \bibinfo{author}{\bibfnamefont{C.}~\bibnamefont{Hempel}},
  \bibinfo{author}{\bibfnamefont{E.}~\bibnamefont{Martinez}},
  \bibinfo{author}{\bibfnamefont{W.}~\bibnamefont{D{\"u}r}},
  \bibinfo{author}{\bibfnamefont{H.}~\bibnamefont{Briegel}},
  \bibinfo{author}{\bibfnamefont{R.}~\bibnamefont{Blatt}}, \bibnamefont{and}
  \bibinfo{author}{\bibfnamefont{C.}~\bibnamefont{Roos}},
  \bibinfo{journal}{Physical review letters} \textbf{\bibinfo{volume}{111}},
  \bibinfo{pages}{210501} (\bibinfo{year}{2013}).

\end{thebibliography}

\end{document}